\documentclass[fleqn,usenatbib]{mnras}

\usepackage[T1]{fontenc}

\DeclareRobustCommand{\VAN}[3]{#2}
\let\VANthebibliography\thebibliography
\def\thebibliography{\DeclareRobustCommand{\VAN}[3]{##3}\VANthebibliography}

\usepackage{graphicx}	
\usepackage{amsmath}	
\usepackage{amssymb}	
\usepackage{nicefrac}
\usepackage{caption}
\usepackage{subcaption}
\usepackage{dcolumn}
\usepackage{array,booktabs}
\usepackage{xcolor,soul}
\sethlcolor{yellow}

\def\vhrulefill{\leavevmode\leaders\hrule height 0.7ex depth \dimexpr0.4pt-0.7ex\hfill\kern0pt}

\newcommand{\msun}{\ensuremath{\mbox{M}_{\odot}}}
\newcommand{\lsun}{\ensuremath{\mbox{L}_{\odot}}}
\newcommand{\rsun}{\ensuremath{\mbox{R}_{\odot}}}

\newcommand{\zaql}{\ensuremath{\zeta~\mbox{Aql~A}}}
\newcommand{\zAql}{\ensuremath{\zeta~\mbox{Aql~A}}}
\newcommand{\zAqlx}{\ensuremath{\zeta~\mbox{Aql}}}
\newcommand{\mum}{\ensuremath{\mu\mbox{m}}}
\newcommand{\kms}{\ensuremath{\mbox{km s}^{-1}}}
\newcommand{\omcrit}{\ensuremath{\omega_{\rm c}}}

\newcommand{\omegak}{\ensuremath{\omega_{\rm k}}}
\newcommand{\omegae}{\ensuremath{\omega_{\rm e}}}
\newcommand{\omomc}{\ensuremath{\omega_{\rm e}/\omega_{\rm c}}}
\newcommand{\prot}{\ensuremath{P_{\rm rot}}}
\newcommand{\prote}{\ensuremath{P_{\rm rot(e)}}}
\newcommand{\pfot}{\ensuremath{P_{\rm phot}}}

\newcommand{\teff}{\ensuremath{T_{\rm eff}}}
\newcommand{\teffl}{\ensuremath{T_{\rm eff}^{\ell}}}

\newcommand\T{\rule{0pt}{2.6ex}}       
\newcommand\B{\rule[-1.2ex]{0pt}{0pt}} 

\newcommand{\rpole}{\ensuremath{R_{\rm p}}}
\newcommand{\gpole}{\ensuremath{g_{\rm p}}}
\newcommand{\tpole}{\ensuremath{T_{\rm p}}}
\newcommand{\xcore}{\ensuremath{\text{X}_{\rm c}}}

\newcommand{\req}{\ensuremath{R_{\rm e}}}
\newcommand{\teq}{\ensuremath{T_{\rm e}}}
\newcommand{\thtbar}{\ensuremath{\overline{\theta}}}

\newcommand{\loggp}{\ensuremath{\log(g_{\rm p})}}
\newcommand{\logge}{\ensuremath{\log(g_{\rm e})}}

\newcommand{\logL}{\ensuremath{\log(L/\lsun)}}

\newcommand{\vesini}{\ensuremath{v_{\rm e}\sin{i}}}
\newcommand{\ve}{\ensuremath{v_{\rm e}}}
\newcommand{\veq}{\ensuremath{v_{\rm e}}}
\newcommand{\sini}{\ensuremath{\sin{i}}}

\newcommand{\SiII}{\ensuremath{\mbox{Si\,\sc{ii}}}}

\newcommand{\hipparcos}{\textit{Hipparcos}}
\newcommand{\gaia}{\textit{Gaia}}
\newcommand{\ester}{\textsc{ester}}
\newcommand{\Ester}{\textsc{Ester}}

\newcommand{\0}{\phantom{0}}

\title[The rapid rotator $\zeta$ Aql]{A study of the rapid rotator \zAqlx:  differential surface rotation?}

\author[I.D. Howarth et al.]{Ian D. Howarth,$^{1}$\thanks{e-mail: i.howarth@ucl.ac.uk}
Jeremy Bailey,$^{2}$ Daniel V. Cotton,$^{3,4}$ and Lucyna Kedziora-Chudczer$^{5}$
\\
$^{1}$University College London, Gower Street, London WC1E 6BT, UK.\\
$^{2}$School of Physics, University of New South Wales, Sydney, NSW 2052, Australia\\
$^{3}$Monterey Institute for Research in Astronomy, 200 Eighth Street, Marina, CA, 93933, USA.\\
$^{4}$Western Sydney University, Locked Bag 1797, Penrith-South DC, NSW 1797, Australia.\\
$^{5}$ Centre for Astrophysics, University of Southern Queensland, Toowoomba, QLD 4350, Australia.}
\date{Accepted 2023 January 10. Received 2023 January 9; in original form 2022 December 12}

\pubyear{2022}

\begin{document}
\label{firstpage}
\pagerange{\pageref{firstpage}--\pageref{lastpage}}
\maketitle

\begin{abstract}
We report new, extremely precise, photo\-polarimetry of the rapidly-rotating A0 main-sequence star \zAqlx,
covering the wavelength range $\sim$400--900nm, which reveals a rotationally-induced signal.   We model the polarimetry, together with the flux distribution and line profiles, in the framework of Roche geometry with $\omega$-model gravity darkening, to establish the stellar parameters.   An additional constraint is provided by \textit{TESS} photo\-metry, which shows variability with a period, \pfot, of 11.1~hr.  Modelling based on solid-body surface rotation 
gives rotation periods, \prot, that are in only marginal agreement with this value.
We compute new \ester\ stellar-structure models to predict horizontal surface velocity fields, which  depart from solid-body rotation at only the $\sim$2\%\ level (consistent with a reasonably strong empirical upper limit on differential rotation derived from the line-profile analysis).   These  models bring the equatorial rotation period, \prote, into agreement with \pfot, without requiring any `fine tuning' (for the \gaia\ parallax).  We confirm that surface abundances are significantly subsolar ($\mbox{[M/H]} \simeq -0.5$).
The star's basic parameters are established with reasonably good precision: 
\mbox{$M = 2.53\pm0.16\,\msun$}, \mbox{$\logL = 1.72\pm0.02$,}
\mbox{$\rpole = 2.21\pm 0.02\,\rsun$}, \mbox{$\teff = 9693 \pm 50$~K}, \mbox{$i = 85{^{+5}_{-7}}^\circ$}, and \mbox{$\omomc = 0.95\pm0.02$}.
Comparison with single-star, solar-abundance stellar-evolution models incorporating rotational effects shows excellent agreement
(but somewhat poorer agreement for models at $\mbox{[M/H]} \simeq -0.4$).
\end{abstract}


\vspace*{-36pt}



\begin{table}
\caption{Selected basic observational data}
\centering
\tabcolsep 1 pt
\begin{tabular}{lllll}
	\toprule
 Parameter&&Value&\;&\phantom{* }Source\\
	\midrule
Spectral type \T&& A0\;IV--Vnn &&\citet{gray03}\\
Parallax    &&$39.28 \pm 0.16$ mas&&\citet{vanLeeuwen07}\\
            &&$38.23 \pm 0.35$ mas&&\citet{gaia3a, gaia3b}\\
\vesini\ &&306$^{+20}_{-5}$~\kms&&Section~\ref{sec:vsini}, Appendix~\ref{sec:rotv}\\
$V$&&2.99&&\citet{johnson66}\\
  &&2.98&&\citet{hoggkvist69}\\
$E(B-V)$&&0\fm005 &&Section~\ref{sec:ispol}\\
$f$(123.5--321nm) &&$4.76\times10^{-7}$ &&Section~\ref{sec:teff}\\
                  &&\multicolumn{1}{r}{erg cm$^{-2}$ s$^{-1}$}&&\B\\

\multicolumn{3}{l}{\textit{Inter\-ferometric results:}}\\
$\thtbar$ \T && $0.895 \pm 0.017$ mas&&\citet{boyajian12}\\
    && $0.961 \pm 0.007$ mas&&\citet{peterson06}\\
\omomc && $0.990 \pm 0.005$ &&(but see Section~\ref{sec:disco})\\
$i$  && $90_{-5}^{+0\,\circ}$ &&\phantom{Peterson}{"}\\
$\theta_*$  && $45 \pm 5^\circ$ &&\phantom{Peterson}{"}\\
\bottomrule
\end{tabular}
\begin{flushleft}
$\thtbar$ is the geometric mean of the  major- and minor-axis limb-darkened angular diameters;  $\theta_*$ is the position angle of the stellar rotation axis, which is at an angle $i$ to the line of sight.\newline

\end{flushleft}
\label{tbl:basic}
\end{table}

\section{Introduction}

The discovery of rotationally-induced, wavelength-dependent linear polarization in
Regulus 
($\alpha$~Leo; \citealt{cotton17}) unveiled a new tool for investigating the properties of rapid rotators, opening the possibility of reasonably precise determinations of mass (and other parameters) in single stars.    Subsequent studies have shown that high-quality photo\-polarimetry can afford power\-ful tests of 
stellar-atmosphere physics and of 
evolutionary models incorporating rotation
(\citealt{bailey20b}; \citealt{lewis22}).  

Here we examine new results for \zAqlx\ (HD~177724, HR~7235, `Okab'), to explore further the diagnostic potential of very precise photo\-polarimetry.
Selected basic data for the star are assembled in Table~\ref{tbl:basic}, including  some of the observational material reviewed in Section~\ref{sec:obsmain}.   Section~\ref{sec:models} outlines our modelling procedures.  Results are given in Section~\ref{sec:results}, which examines the question of possible differential surface rotation.    Section~\ref{sec:disco} confronts the inferred stellar parameters with stellar-evolution models.   The determination of \vesini, and other aspects of the line-profile analysis, are relegated to Appendix~\ref{sec:rotv}.

\section {Observations}
\label{sec:obsmain}

\begin{figure}
\includegraphics[width=\columnwidth]{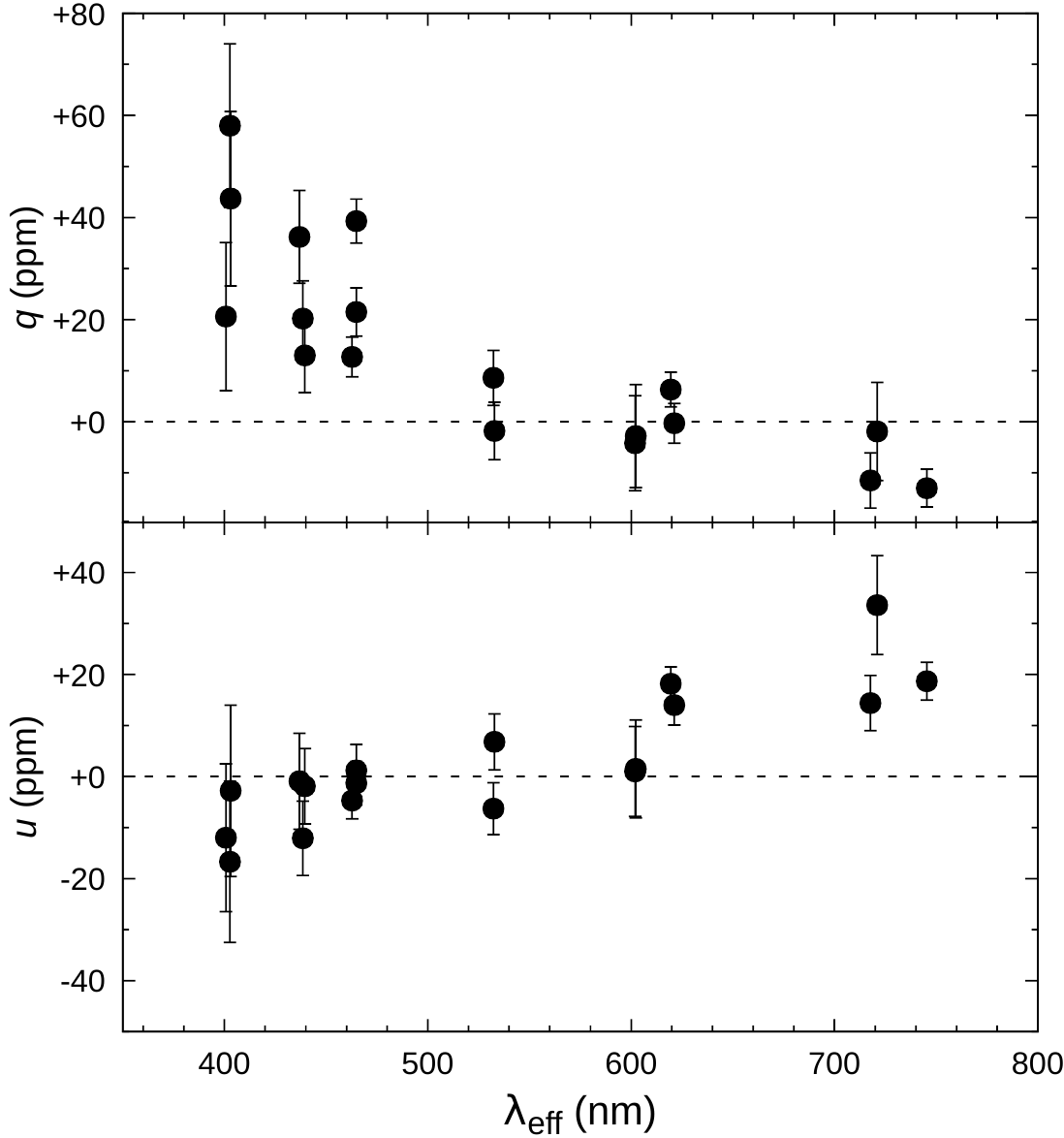}
\caption{Observed photopolarimetry of $\zeta$~Aql.}
 \label{fig:obs}
\end{figure}

\begin{figure}
\includegraphics[width=\columnwidth]{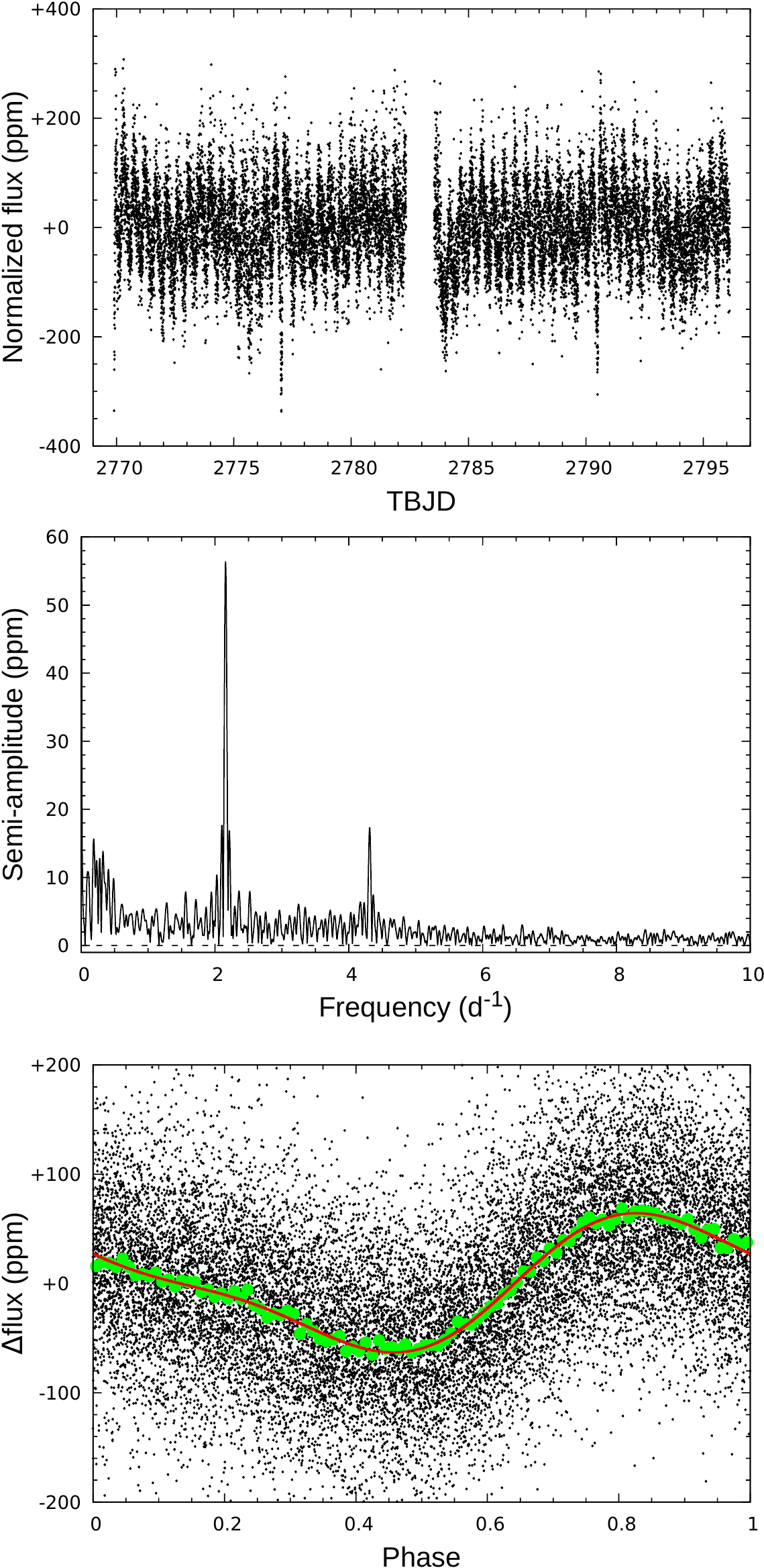}
\caption{\textit{TESS} light-curve of \zAqlx. Top panel: \mbox{PDCSAP}  flux
(`pre-search data conditioning simple aperture', normalized simply by dividing the mean and subtracting unity) as a function of \textit{TESS} Barycentric Julian Date ($=\text{BJD} - 2\,457\,000.0$,
$\simeq\text{MJD} - 56\,999.5$).   Middle panel: generalized Lomb-Scargle periodogram.   Bottom panel:  photometry phase-folded at \mbox{$\pfot = 11.118$~hr} (with phase zero arbitrarily defined as time of first observation);  green dots are data averaged in 0.01 phase bins, and the red line is the two-component Fourier model.}
\vspace{-1mm}  
\label{fig:tesslc}
\end{figure}

\subsection{Photopolarimetry}
\label{sec:ppol}

The new observations reported here were obtained using HIPPI, the HIgh-Precision Polarimetric Instrument, and its successor, HIPPI-2;
the design and operation of these instruments are described by \citet{bailey15,bailey20a}. 
We also made use of one previously published observation \citep{bailey10}, obtained using PlanetPol \citep{hough06}.

Details of observing runs and instrument configurations are given in 
Appendix~\ref{sec:obsruns} (Table~\ref{tab:runs}), and
the individual observations of \zAqlx\ are listed in Table~\ref{tbl:obs}. As well as shot noise, the uncertainties reported therein on the normalized Stokes parameters
$q, u$, and on the polarization $p$, include (in quadrature)
a wavelength-dependent positioning error which results from inhomogeneities across the face of the ferro-electric liquid-crystal modulator, and which sets the accuracy limit for the instruments. For HIPPI-2 this limit ranges from 1.1 parts per million (ppm) for the reddest passbands, through 2.5~ppm in $g^\prime$, to 13.7~ppm in the 425SP filter -- similar to, but slightly better than, the corresponding figures for HIPPI \citep{bailey20a}.
Position-angle (PA) calibration was performed using highly-polarized standard stars; typical PA zero-point uncertainties are
$\sim{1}^\circ$ (i.e., are smaller than the statistical errors on our measurements).

Results are plotted in Fig.~\ref{fig:obs}; the scatter in the observations slightly exceeds 
the quoted formal errors (though is still small compared to results from traditional polarimeters),
as was also found in our study of $\theta$~Sco \citep{lewis22}.  Although  intrinsic low-level polarization variability cannot be ruled out, the scatter is consistent with a combination of
instrument-configuration changes and imperfect characterization of low-polarization standard stars (used to correct for polarization arising in the telescope optics), which 
can lead to zero-point drifts of up to $\sim$10~ppm between runs \citep{bailey21}. These factors are accommodated in our modelling by use of bootstrapping in the error analysis (Section~\ref{sec:results}).

\begin{table*}
\caption{Photopolarimetry of \zAqlx, sorted by passband effective wavelength, $\lambda_{\rm eff}$.
Run identifiers correspond to entries in Table~\ref{tab:runs}, where further technical details are
given, and reflect the year and month of each observing campaign. Dwell times include observing overheads, and so exceed actual integration times (`Int.').
`Det.' indicates whether a B(lue) or R(ed) photo\-multiplier tube was used as detector; `Eff.' is the modulator polarization efficiency. The final four columns give the normalized Stokes parameters, $q$, $u$;  the polarization, $p$;  and the observed polarization position angle, $\theta_{\rm p}$ (where $q = Q/I = p \cos 2\theta_{\rm p}$, $u = U/I = p\sin 2\theta_{\rm p}$, and 
$0^\circ \le \theta_{\rm p} < 180^\circ$).}
\label{tbl:obs}
\tabcolsep 4 pt
\begin{tabular}{lccrcccrrrrrr}
\toprule
 \;Run ID & \multicolumn{1}{c}{MJD}       & Dwell & Int. & Filter & Det. & $\lambda_{\rm eff}$ & Eff. & \multicolumn{1}{c}{$q$} & \multicolumn{1}{c}{$u$} & \multicolumn{1}{c}{$p$} & \multicolumn{1}{c}{$\theta_{\rm p}$}\\
        & (mid-dwell) & \multicolumn{1}{c}{(s)} & \multicolumn{1}{c}{(s)} & & & \multicolumn{1}{c}{(nm)} & \multicolumn{1}{c}{(\%)} & \multicolumn{1}{c}{(ppm)} & \multicolumn{1}{c}{(ppm)} & \multicolumn{1}{c}{(ppm)} & \multicolumn{1}{c}{($^\circ$)}\\
\midrule
2017\_08 & 57977.530 
 & 3458 & 2560 & 425SP & B & 400.7 & 52.2 &   20.6 $\pm$  14.5 &  $-$12.0 $\pm$  14.5 &   23.8 $\pm$  14.5 &  164.9 $\pm$  21.6 \\
2018\_07 & 58318.566 
 & 1426 & 960 & 425SP & B & 402.7 & 38.3 &   58.0 $\pm$  16.0 &  $-$16.7 $\pm$  15.8 &   60.4 $\pm$  15.9 &  172.0 $\pm$   \07.7 \\
2018\_07 & 58315.473 
 & 1077 & 640 & 425SP & B & 403.0 & 38.5 &   43.7 $\pm$  17.1 &   $-$2.8 $\pm$  16.8 &   43.8 $\pm$  17.0 &  178.2 $\pm$  12.5 \\
2017\_08 & 57977.569 
 & 1870 & 640 & 500SP & B & 436.9 & 75.4 &   36.2 $\pm$   \09.1 &   $-$0.9 $\pm$   \09.4 &   36.2 $\pm$   \09.3 &  179.3 $\pm$   \07.5 \\
2018\_07 & 58318.525 
 & 1058 & 640 & 500SP & B & 438.5 & 67.1 &   20.2 $\pm$   \07.4 &  $-$12.1 $\pm$   \07.3 &   23.5 $\pm$   \07.4 &  164.5 $\pm$   \09.4 \\
2018\_07 & 58315.459 
 & 1132 & 640 & 500SP & B & 439.5 & 67.9 &   13.0 $\pm$   \07.3 &   $-$1.9 $\pm$   \07.4 &   13.1 $\pm$   \07.4 &  175.8 $\pm$  19.9 \\
2018\_07 &58318.511 
 & 1161 & 640 & $g^{\prime}$ & B & 462.8 & 79.6 &   12.7 $\pm$   \03.9 &   $-$4.7 $\pm$   \03.6 &   13.5 $\pm$   \03.8 &  169.8 $\pm$   \08.1 \\
2015\_10 & 57314.382 
 & 1238 & 640 & $g^{\prime}$ & B & 464.8 & 89.4 &   39.3 $\pm$   \04.3 &   $-$1.3 $\pm$   \04.3 &   39.3 $\pm$   \04.3 &  179.1 $\pm$   \03.2 \\
2017\_08 & 57977.569 
 & 2610 & 640 & $g^{\prime}$ & B & 464.8 & 86.9 &   21.5 $\pm$   \04.7 &    1.2 $\pm$   \05.1 &   21.5 $\pm$   \04.9 &    1.6 $\pm$   \06.7 \\
2018\_07 & 58318.551 
 & 1067 & 640 & $V$ & B & 532.3 & 95.6 &    8.6 $\pm$   \05.4 &   $-$6.3 $\pm$   \05.1 &   10.7 $\pm$   \05.3 &  161.9 $\pm$  17.1 \\
2018\_07 & 58315.445 
 & 1039 & 640 & $V$ & B & 532.8 & 95.6 &   $-1.8$ $\pm$   \05.6 &    6.8 $\pm$   \05.5 &    7.0 $\pm$   \05.6 &   52.4 $\pm$  27.3 \\
2018\_07 & 58318.538 
& 1066 & 640 & $r^{\prime}$ & B & 601.9 & 86.8 &   $-$4.2 $\pm$   \09.3 &    1.0 $\pm$   \08.8 &    4.3 $\pm$   \09.0 &   83.3 $\pm$  42.1 \\
2018\_07 & 58315.432 
& 1033 & 640 & $r^{\prime}$ & B & 602.3 & 86.8 &   $-$2.8 $\pm$  10.1 &    1.5 $\pm$   \09.6 &    3.2 $\pm$   \09.8 &   75.9 $\pm$  45.3 \\
2017\_08 & 57973.533 
 & 3551 & 2560 & $r^{\prime}$ & R & 619.5 & 82.6 &    6.3 $\pm$   \03.4 &   18.2 $\pm$   \03.3 &   19.3 $\pm$   \03.4 &   35.5 $\pm$   \05.0 \\
2018\_07 & 58322.569 
 & 2508 & 960 & $r^{\prime}$ & R & 621.2 & 83.1 &   $-$0.3 $\pm$   \03.9 &   14.0 $\pm$   \03.9 &   14.0 $\pm$   \03.9 &   45.6 $\pm$   \08.1 \\
2017\_08 & 57973.492 
 & 3404 & 2560 & 650LP & R & 717.7 & 65.9 &  $-$11.5 $\pm$   \05.4 &   14.4 $\pm$   \05.4 &   18.4 $\pm$   \05.4 &   64.3 $\pm$  \08.7 \\
2018\_07 & 58322.602 
 & 1439 & 640 & 650LP & R & 721.0 & 65.0 &   $-1.9$ $\pm$   \09.6 &   33.6 $\pm$   \09.7 &   33.7 $\pm$   \09.6 &   46.6 $\pm$   \08.4 \\
2005\_04$^a$ & 53494.216 &   1830 & 1440 & BRB & APD & 745.4 & 92.3 &  $-$13.0 $\pm$   \03.7 &   18.7 $\pm$   \03.7 &   22.8 $\pm$   \03.7 &   62.4 $\pm$  \04.7 \\

\bottomrule
\end{tabular}
\begin{flushleft}
($a$) The PlanetPol observation comes from \citet{bailey10}.  
\end{flushleft}
\end{table*}

\begin{figure*}
\includegraphics[width=\textwidth]{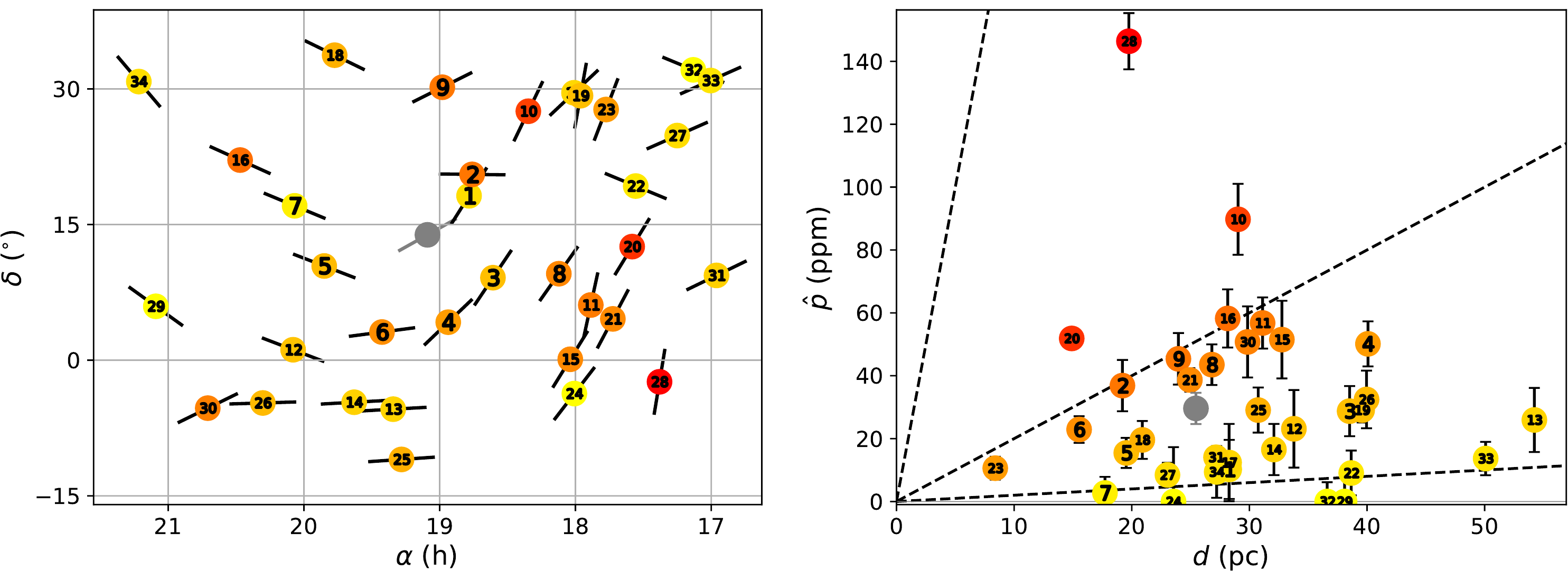}
\caption{(Left) map and (right) bias-corrected polarization, $\hat{p}, $ vs.\ distance $d$ for inter\-stellar control stars within 35$^\circ$ and 30~pc of $\zeta$~Aql\ (which is indicated by the grey data-points;  
in the right-hand panel, this is the prediction of the inter\-stellar model of \citealt{cotton17b}).\newline
Distances and co-ordinates were obtained from SIMBAD (mostly \gaia\ DR3 values, with a  handful of \hipparcos\ results), and  polarization measure\-ments from \citet{bailey10}, \citet{marshall16}, \citet{piirola20}, and \citet{bailey20b}, with two additional values from \citeauthor{marshallPrep} (in prep.). 
Black pseudo-vectors on the map points indicate the position angles (but not the magnitudes) of the inter\-stellar polarizations.\newline 
For the $d$--$\hat{p}$ plot, observed polarizations $p$ were debiased using the method of \citet{wardle74}, as first discussed by \citet{serkowski58}: $\hat{p} = \left({p^2-\sigma_p^2}\right)^{1/2}$ for $p>\sigma_p$, or $\hat{p} = 0$ otherwise (which corrects for $p$ being positive definite). We then transformed the multi-wavelength results to a standardized effective wavelength of 450~nm  by adopting a Serkowski law (eqtn.~\ref{eq:serk}) with $\lambda_{\max} = 470$~nm, appropriate for stars within the Local Hot Bubble \citep{marshall16, cotton19b}, and $K$ set by eqtn.~\eqref{eq:kserk}. Stars are colour-coded in terms of increasing $\hat{p}/d$ (yellow$\rightarrow$red) and numbered in order of increasing angular separation from $\zeta$~Aql: \newline1,~HD~173880; 2,~HD~173667; 3,~HD~171802; 4,~HD~175638; 5,~HD~187691; 6,~HD~182640; 7,~HD~190406; 8,~HD~165777; 9,~HD~176337; 10, HD~168874; 11,~HD~162917; 12,~HD~190412; 13,~HD~181391; 14,~HD~185124; 15,~HD~164651; 16,~HD~195034; 17,~HD~164595; 18,~HD~187013; 19,~HD~163993; 20,~HD~159561; 21,~HD~161096; 22,~HD~159332; 23,~HD~161797; 24,~HD~164259; 25,~HD~180409; 26,~HD~193017; 27,~HD~156164; 28,~HD~157347; 29,~HD~200790; 30,~HD~197210; 31,~HD~153210; 32,~HD~155060; 33,~HD~153808; 34,~HD~202108. 
\newline
Dashed lines, given as guides in the right-hand panel, correspond to $\hat{p}/d$ values of 0.2, 2.0, and 20.0~ppm~pc$^{-1}$.}
 \label{fig:is_map}
\end{figure*}

\subsection{\textit{TESS} photometry}
\label{sec:tess}

The \textit{TESS} satellite \citep{ricker15}  observed  \zAqlx\ in sector~54 (2022 July--August). The light-curve is shown in Fig.~\ref{fig:tesslc};  a periodic signal is immediately evident.  A generalized Lomb-Scargle periodogram 
(\citealt{zechmeister09}; \citealt{ferraz81}) yields a fundamental frequency and corresponding semi-amplitude of \\*
$\phantom{X} \nu_0= 2.1587 (8)$ d$^{-1}$ \qquad [$\pfot = 11.118 (4)$ hr]\\*
$\phantom{X} a_0=56.5 (25)$ ppm\\
where parenthesized values are 1-$\sigma$ uncertainties on the least significant digits, generated from Monte-Carlo simulations using a residual-permutation, or `prayer beads', algorithm (adopted because the residuals are strongly correlated as a result of lower-frequency drifting).   The signal appears to be rather simple;  power at the first harmonic accounts almost entirely for departures from a pure sinusoid (Fig.~\ref{fig:tesslc}).

\subsection{Ancillary observational data}

Some additional observational material is required for our analysis;  furthermore, given the level of precision of the \mbox{polarimetry}, consideration needs to be given to potential contaminating sources (whether stellar or circumstellar).

\subsubsection{Parallax}

Parallaxes, required principally to establish the stellar radius, are available from both the \hipparcos\ and \gaia\ missions
(\citealt{vanLeeuwen07}; EDR3, \citealt{gaia3a,gaia3b}).   The \hipparcos\ result is
$1.05\pm0.38$~mas larger than the EDR3 value (Table~\ref{tbl:basic}), a 2.6-$\sigma$ difference.   Although this difference is of little consequence for most derived parameters (cf.\ Table~\ref{tbl:sensit}), it proves to be of some significance for the interpretation of the photo\-metric period (Section~\ref{sec:pprob}).   We therefore performed calculations for both values (taking the distance, $\sim$26~pc, to be simply the inverse of the parallax\footnote{The adjusted \mbox{\gaia} distances given by \mbox{\citet{bailerjones21}} are only $\sim$0.2$\sigma$ (or $\sim$0.2\%) smaller.  The biasses discussed by \mbox{\citet{Lindegren21}} are not defined for stars as bright as \zAqlx, but are typically at the $\sim$10$\mu$as level.}).

\subsubsection{Inter\-stellar polarization and extinction}
\label{sec:ispol}

\citet{cotton17} developed a model of inter\-stellar polarization from which we expect 
$p(\lambda_{\rm max}) \simeq 30$~ppm for \zAqlx, where $\lambda_{\rm max}$ is  the wavelength of maximum linear polar\-ization (typically $\sim$0.5\mum).    Observations of stars over a range of distances in the general direction of \zaql\ support this estimate, with an upper limit of $\sim$50--60~ppm (Fig.~\ref{fig:is_map}).
The inter\-stellar polarization is directly estimated as part of the modelling (Section~\ref{sec:modover}), and yields $p(\lambda_{\max})\sim$17--24~ppm (in position angle $\theta_{\rm i} = 55 \pm 2^\circ$), in good accord with the 
\citeauthor{cotton17} model.

Such small polarizations imply very little foreground dust and inter\-stellar reddening.
\citet{serkowski75} found \mbox{$E(B-V) \gtrsim p/9$\%} for nearby stars, suggesting
a barely non-zero extinction.   For the purposes of correcting the  observed flux distribution
we adopt \mbox{$E(B-V) = 0\fm005$} as a suitably small, if arbitrary, round-number estimate;  our results are very insensitive to the exact value (Table~\ref{tbl:sensit}).

\subsubsection{Projected rotation velocity}
\label{sec:vsini}

The projected equatorial rotation velocity, \vesini, provides an important constraint on the modelling.    Our examination of \vesini\ is described in detail in Appendix~\ref{sec:rotv}.   There is a modest dependence of the inferred value on \omomc, 
the ratio of the equatorial angular
velocity to the critical value at which the Newtonian gravitational
force is matched by the centrifugal force,
\begin{align}
\omcrit = \sqrt{
{(G M)}/ {(1.5 \rpole)^3}
}
\label{eq:vcrit}
\end{align}
(for a star of mass $M$ and polar radius \rpole).

There is, additionally, some sensitivity to the surface-rotation profile (i.e., the variation, or otherwise, of $\omega$ with colatitude $\theta$).   For example, models generated with the
\ester\ stellar-structure code, discussed in Appendix~\ref{sec:estermod}, have a differential-rotation profile that results in \vesini\ values 
$\sim$4~\kms\ smaller than does solid-body surface rotation.

Our modelling takes these factors fully into account;   the overall range of acceptable \vesini\ values is $\sim$300--325~\kms, with $\vesini = 306$~\kms\ for our final preferred model (Section~\ref{sec:results}).

\subsubsection{Companion stars}
\label{sec:compstars}
The Washington Double Star Catalog (WDS;  \citealt{mason01}) lists four visual companions to \zAql\ (=WDS~J19054+1352A);  of these, only the B~component, at separation $\rho=7\arcsec$ \citep{deRosa14,gaia3a}, is close enough to potentially affect the observations discussed in this paper.  (It is also the only physical companion, according to \gaia\ astrometry.)  

The B component was discovered by \citet{Burnham74}, who described it as ``not fainter than{\ldots}11~mag''.
\citet{Wallenquist47} reported a visual\footnote{\citeauthor{Wallenquist47} used a wedge photo\-meter;  therefore, although the observation was visual, it is nevertheless a measure\-ment, not merely an estimate.}   
magnitude difference of $\Delta{v} = 8\fm45$, while  $\Delta{G}=7\fm85$  \citep{gaia3a} and $\Delta{K} = 4\fm87$ \citep{deRosa14}.     The colours and absolute magnitudes are consistent with an \mbox{early-M} dwarf companion, which would contribute <1\%\ of the flux at $\lambda$<1$\mum$ (<0.1\% at $\lambda$<0.6$\mum$), rising to $\sim$3\%\ only for $\lambda \gtrsim 4\mum$.
The B component is therefore of no importance for the observations and analysis reported here.

As pointed out to us by our referee, the
\mbox{\gaia} 
image parameters can provide additional information on potential close companions.   The Renormalised Unit Weight Error (RUWE) for the 
\mbox{\zAql} astrometric solution is 2.5, which initially appears to be rather large compared to the value of $\sim$1 expected for well-behaved solutions of single stars.  
However, we find that this is value is actually typical of very bright stars;  the 548 stars in DR3 with $G \le 4.0$ have a median RUWE of 2.73.   
The \texttt{ipd\_gof\_harmonic\_amplitude} parameter is a measure of image asymmetry;  its value of 0.09 is fully consistent with a circular image \mbox{\citep{fabricius21}}.
\mbox{\citet{peterson06}} also imply that companions 
with $\Delta{R}\lesssim 8$ within 0\farcs5 of \mbox{\zAqlx} are ruled out by their interferometric observations.

\subsubsection{Is \zAql\ a spectroscopic binary?}
\label{sec:notsb}

 At the time of writing, the WDS carries an unattributed note that ``A is a spectroscopic binary''.    We have been unable to find any documented source of that report.    We therefore examined the sixty-three good-quality, high-resolution spectra discussed in Appendix~\ref{sec:rotv}, obtained between 2005 May and 2014 June, which sample timescales of minutes, hours, days, and years.  Simple visual inspection revealed no obvious line-profile or radial-velocity variations, and
 cross-correlation velocity measure\-ments of  the Ca\;{\sc ii}~$K$ line yield an r.m.s. dispersion of only 3.6~\kms\ (cp.\ the resolution element, $\sim$4.6~\kms, and \vesini, $\sim$300~\kms).   We proceed on the assumption that if \zAql\ is indeed a spectroscopic binary, then that is of no consequence for our analysis.

\subsubsection{Exozodiacal-dust emission}
\label{sec:exoz}
\citet{absil08} reported a $K$-band excess of $(1.69\pm0.27)$\%\ of the photo\-spheric flux,\footnote{The 
similarity of the implied $\Delta{K}$, 4\fm4, to that of the B component must be coincidental; the field of view of the instrument used to detect the excess is only 0\farcs8 (fwhm).   However, \citeauthor{absil08}'s   discussion of the infra-red photo\-metric results will be compromised by their neglect of the companion star's flux contribution.}  based on differences between short-baseline inter\-ferometric visibilities 
observed with CHARA and those predicted from a colour-based surface-brightness estimate of the angular diameter.
Using a different detector (though the same instrument and method\-ology),
\citet{nunez17} obtained a $\Delta{K}$ of
$(1.23 \pm 0.38)$\%, attributing this excess to hot ($\sim$1000~K) exozodiacal dust.

The flux contribution from dust emission of this nature is again too small to have any direct consequences for our study.   In principle, if aligned grains were involved,  they might influence the photo\-polarimetry;  however, current evidence suggests that the grains responsible for exo\-zodiacal emission are too small to produce polarization at optical wavelengths
\citep{marshall16}.

\subsubsection{Inter\-ferometry}

Finally, long-baseline optical inter\-ferometry can provide a useful check on our results.
We found two reports in the literature:  \citet{peterson06} 
briefly summarize an otherwise  unpublished detailed analysis of NPOI measure\-ments,
while \citet{boyajian12} give a mean angular diameter from CHARA data.   Results are included in Table~\ref{tbl:basic};  the angular diameters from the two studies differ by $\sim$7\%\ ($\sim$3.5$\sigma$).  Since the NPOI angular diameter ($\sim$$VR$ passband) is larger than 
the CHARA value  ($\sim$$K$), the discrepancy cannot be attributed to the possible exozodiacal-dust emission
discussed in Section~\ref{sec:exoz}.


\section{Modelling} 
\label{sec:models}

The principal motivation for our  modelling is to determine values for basic stellar parameters from the photo\-polarimetry.  We conducted our analysis in the framework of standard Roche geometry (e.g., \citealt{collins63}), with $\omega$-model gravity darkening \citep{espinosa11}.  

While the photo\-metric variability suggests the possibility of some departure from axial surface-brightness symmetry, it is at a very low level (and in any case,
we have no way of characterizing it in an appropriate manner).   Because our polarimetric observations were taken at arbitrary phases, we do not anticipate systematic effects, and the bootstrap error analysis accommodates any increase in observational scatter that may arise.
\subsection{Overview}
\label{sec:modover}

Photo\-spheric polarization was computed using the code described by \citet{cotton17} and \citet{bailey20b}, which calculates local surface intensities using the {\sc synspec} spectral-synthesis program \citep{hubeny85, hubeny12}, modified for fully polarized radiative transfer using the {\sc vlidort} package \citep{spurr06}.   The underpinning atmosphere models are custom \textsc{Atlas9} line-blanketed LTE calculations \citep{castelli03}.

The model polarization depends principally on four quantities (or equivalent surrogates):   a reference temperature and gravity (e.g., polar values \tpole, \gpole); 
the inclination of the rotation axis to the line of sight, $i$; and the rotation parameter, \omomc.

We can reduce this  four-dimensional parameter dependency to a two-dimensional grid by exploiting complementary observations which independently constrain the temperature and gravity (Section~\ref{sec:gin}), allowing us to compute predicted polarizations as functions of $i$ and $\omomc$ alone (at the self-consistent \tpole, \gpole\ values). 

Final parameter values are then determined by selecting models that best match the observed photo\-polarimetry, as judged by $\chi^2$.   Inter\-stellar polarization has a significantly different wavelength dependence
to rotational effects, and its magnitude and direction can therefore be estimated in parallel with the photospheric-model minimization;   we assume a `Serkowski law' \citep{serkowski73,serkowski75},
\begin{align}
\frac{p(\lambda)}{p(\lambda_{\rm max})} =
\exp\left({-K \ln^2 (\lambda_{\rm max} / \lambda)}\right),
\label{eq:serk}
\end{align}
with
\begin{align}
K = 0.01 + 1.66 \lambda_{\rm{max}}
\label{eq:kserk}
\end{align}
\citep{wilking80, whittet92}.  We fixed $\lambda_{\rm max}$ at 470~nm, a value appropriate to the Local Hot Bubble \citep[][and references therein]{marshall20}, as the inter\-stellar polarization proves to be too small to allow an independent determination to useful accuracy.

\subsection{Reducing the 4-D dependency}
\label{sec:gin}

\subsubsection{Temperature}
\label{sec:teff}

The method underpinning our temperature determinations is closely akin to the Infra-Red Flux Method of \citet{blackwell77}, the basic principle being that  the ratio of the observed fluxes in two suitable regions is a measure of temperature.   Ideally, `suitable' regions should have strongly different temperature dependences (e.g., be on either side of the peak of the flux distribution), and should record a significant part of the total luminosity.   

We used $V$ photo\-metry and the UV flux from  observations made with the \textit{International Ultraviolet Explorer} (\textit{IUE}), as summarized in Table~\ref{tbl:basic}.   We investigated the use of longer-wavelength ($RIJ$) photo\-metry in place of $V$, but found that this did not afford any useful gain in sensitivity (in part because of increasing observational uncertainties).  

We used all available archival \textit{IUE} observations obtained through its spectrographs' large apertures at low resolution (resolving power $R\simeq 350$), which provide the most reliable flux measure\-ments.   The nine short-wavelength ($\sim$115--198~nm) and nine long-wavelength 
($\sim$185--335~nm) flux-calibrated spectra were combined with weights proportional to exposure times, excluding image LWR~7295, which has a notably poor signal:noise ratio.
All fluxes were corrected for the adopted inter\-stellar extinction using a \citet{seaton79} curve, with $A_V/E(B-V) = 3.1$.   The UV flux accounts for $\sim$25--30\%\ of the luminosity, and the $V$~band for $\sim$10\% (cf.\ Fig.~\ref{fig:flxplt}).

\begin{figure*}
\includegraphics[width=0.75\textwidth]{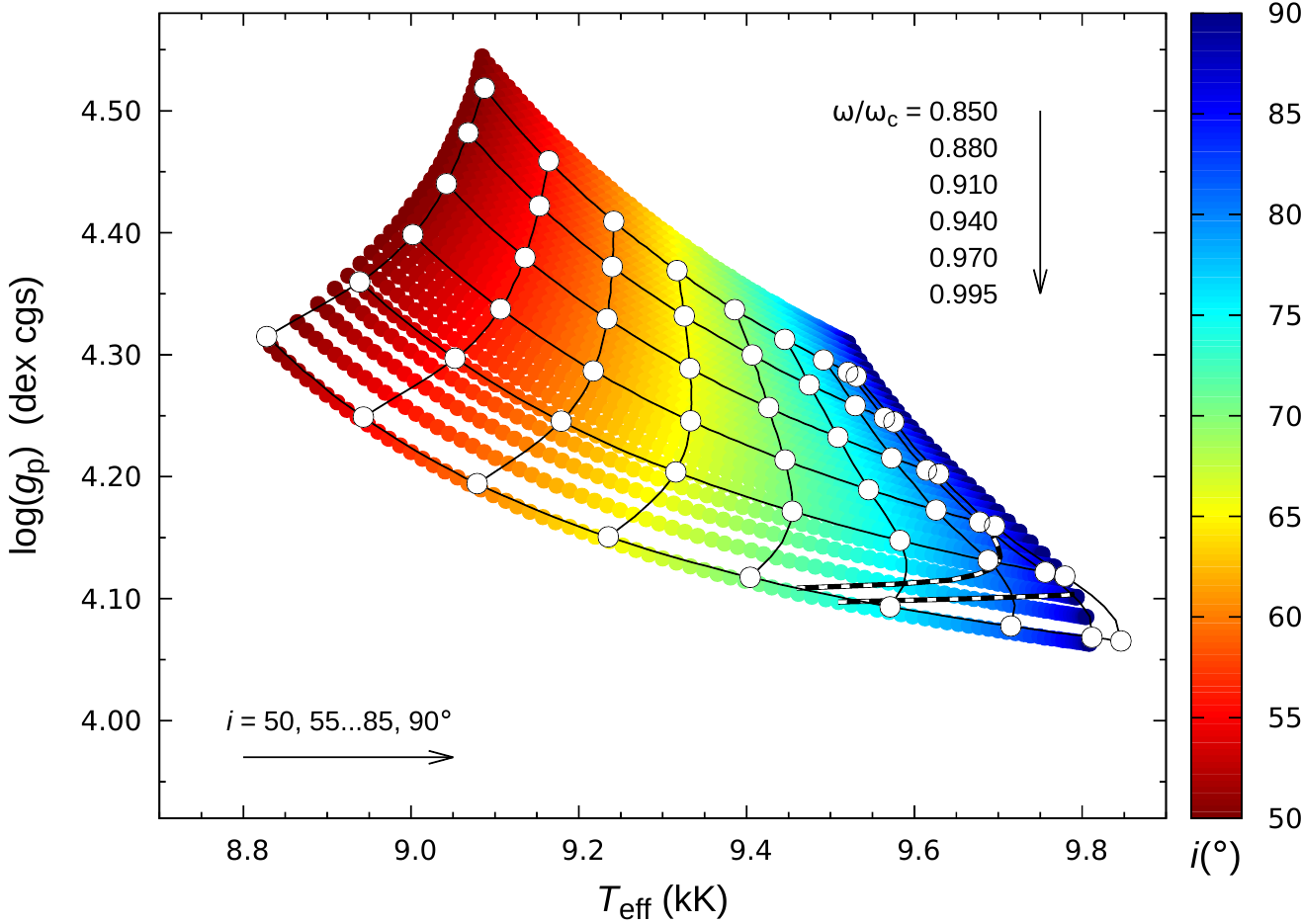}
\caption{Grids of \teff, \loggp\ values that reproduce the observed $V$, UV fluxes as functions of \omomc\ and axial inclination, $i$.   The base grid, colour-coded for inclination, shows results for the full set of models based on solid-body surface rotation and the \hipparcos\ parallax ($i = 50$--90$^\circ$ at 0.5$^\circ$ steps; \omomc\ in the range 0.850--0.995 at steps of 0.005, with \vesini\ from
eqtn.~\ref{eq:voma}).    The sparse grid of connected larger white dots shows a subset of corresponding results for models based on the \gaia\ parallax and \ester\ differential rotation (Appendix~\ref{sec:estermod}), sampled as labelled in the Figure.  The near-horizontal dashed lines are loci of models from each grid which have equatorial rotation periods that match the \textit{TESS} photo\-metric period.}

\label{fig:pltGrid}
\end{figure*}

\subsubsection{Metallicity, radius}

\begin{figure*}
\includegraphics[width=0.75\textwidth]{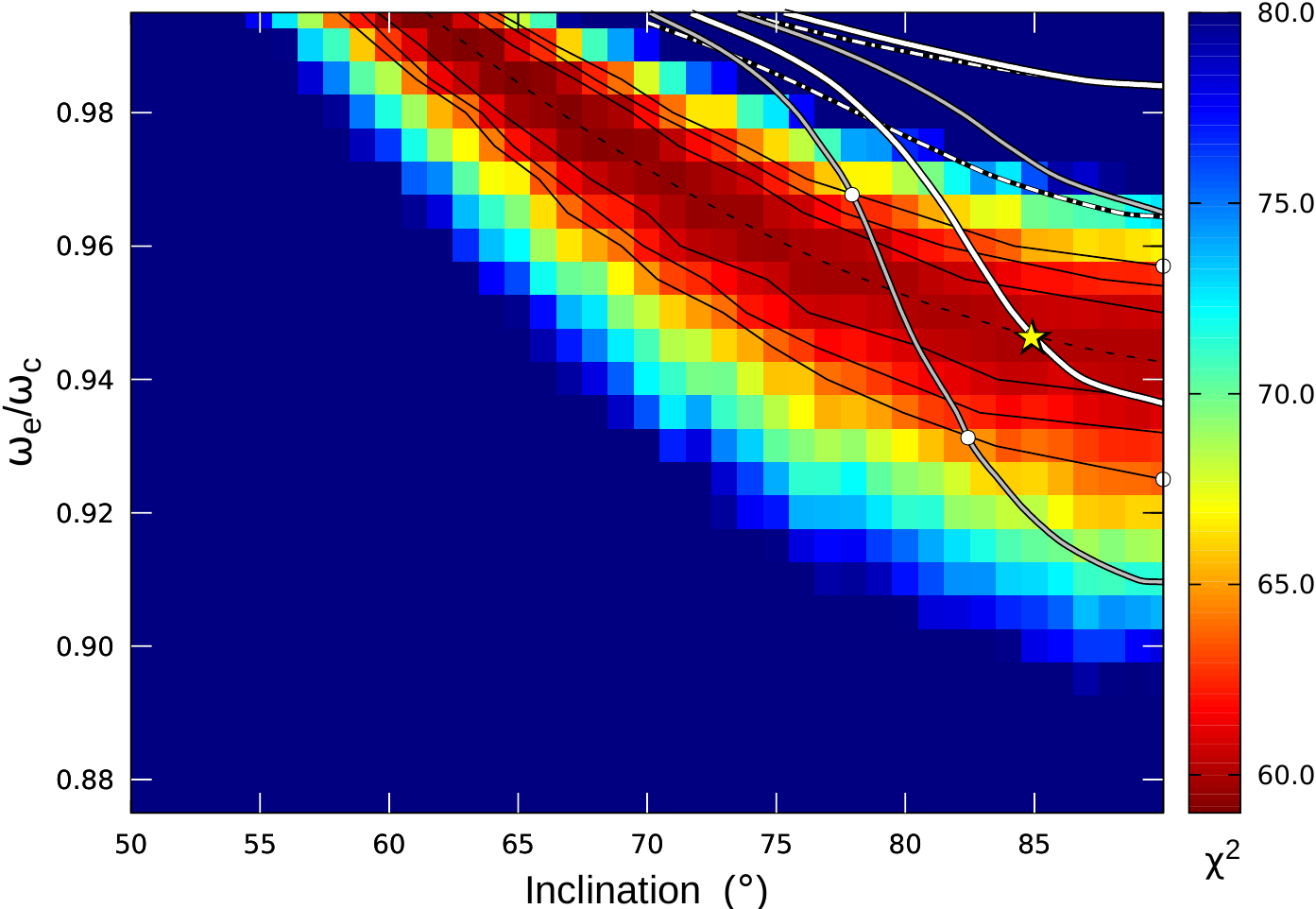}
\caption{Map of $\chi^2$ comparisons of observed and modelled polar\-izations (models based on solid-body surface rotation, the \hipparcos\ parallax, and  $\mbox{[M/H]}=-0.5$).     The dashed black line shows an analytical approximation to the minimum-$\chi^2$ locus (eqtn.~\ref{eq:valley}), with solid black lines corresponding to  1-, 2-, and 3-$\sigma$ confidence intervals on the range of acceptable models (Section~\ref{sec:results}).
Dash-dot lines show the loci of solid-body surface-rotation models with equatorial rotation periods that match the \textit{TESS} photo\-metric period (upper, lower for \hipparcos, \gaia\ parallaxes, respectively). The solid white lines are corresponding results for \ester-model differential rotation.
Grey lines are for \gaia-parallax, \ester-rotation models which have equatorial rotation periods  differing from \pfot\ by $+1/-$1\%\  (upper/lower lines).  White dots mark four of the models listed in Table~\ref{tbl:params2}, with the yellow star corresponding to the adopted `base' model therein.}
\label{fig:JABchi2}
\end{figure*}

Conversion of the observed flux ratio to a temperature is achieved by means of model-atmosphere flux distributions.   This introduces a metallicity dependence, principally through line blanketing, whereby the observed UV flux is reproduced by lower-temperature models at lower metallicity.
\citet{gray03} and \citet{wu11} report $\mbox{[M/H]}=-0.68\pm0.09, -0.52\pm0.16$, respectively, for \zAqlx;  the synthetic spectra described in Appendix~\ref{sec:rotv}  also indicate substantially subsolar metallicity (cf.\ Fig~\ref{fig:rotprof}).   Our analysis is not sensitive to the precise value (Table~\ref{tbl:sensit}); we adopt $\mbox{[M/H]}=-0.5$  as a suitable round-number value (for calculation of both parameter grids and model polarizations).  This results in effective temperatures $\sim$2\%\ lower than would be inferred from solar-abundance models.\footnote{We define the (global) effective temperature for a gravity-darkened star through the stellar luminosity,
\begin{align}
\teff^4 = {{\int{(\teffl)^4\,\text{d}A}}\left/{{{\int{\text{d}A}}}}\right.},
\end{align}
where \teffl\ is the \textit{local} (latitude-dependent) effective temperature and the integrals are over surface area.   The ratio of polar to effective temperatures is solely a function of \omomc\ (for given gravity-darkening and surface-rotation prescriptions); in the case of  $\omega$-model gravity darkening and rigid-body surface rotation,
$\tpole/\teff = 1.09 \rightarrow 1.16$ for $\omomc = 0.85 \rightarrow 0.99$.}

With the temperature established, the observed flux \textit{level} directly yields the angular diameter, which can be converted into a stellar radius given the distance.

\subsubsection{Rotational effects}
\label{sec:rote}

Significant rotation introduces dependences on \omomc\ and $i$ to the modelled fluxes, through gravity darkening and aspect effects.   However, for a given (or assumed) value of $i$, the equatorial rotation velocity, \ve, follows directly from the observed projected rotation velocity, \vesini, thereby establishing $\omega_{\rm e}$ (from the equatorial radius).   For a given (or assumed) value of \omomc, the corresponding mass can then be inferred (from eqtn.~\ref{eq:vcrit}).   Consequently, for any specified $i$, $\omomc$ combination, there is a unique pair of temperature and polar-gravity values (and associated mass and radius) that reproduce the observed fluxes.

To determine those values in practice, we run a series of models to calculate fluxes at 100-K steps in \teff, for specified $i$, \omomc\ pairs
(at given \vesini, $d$, [M/H], and $E(B-V)$).
These models are computed in full, limb- and gravity-darkened Roche geometry, using the
code described by \citet{howarth16}, with \textsc{Atlas9} intensities
from \citet{howarth11}.
At each \teff\ we determine the polar radius that reproduces the observed flux (UV or $V$) using a simple interval-halving algorithm, in order to establish a locus of acceptable values in \teff, \loggp\ space.
The required final result is given by the intersection of the UV and $V$ loci 
(which is well defined for this \teff\ regime, confirming that these passbands are `suitable' in the sense discussed in Section~\ref{sec:teff}).

\section{Results}
\label{sec:results}

Fig.~\ref{fig:pltGrid} shows selected  results of the \teff/\loggp\ modelling described in Section~\ref{sec:gin}.    Our first parameter grid was constructed with models based on 
the \hipparcos\ parallax (as it is more precise than the \gaia\ value) and 
the `Occam's razor' assumption of solid-body surface rotation (consistent with the quite strong upper limit on differential rotation  obtained from the line-profile analysis reported in Section~\ref{sec:rotnum}), using  eqtn.~\eqref{eq:voma} to link
\vesini\ and \omomc.

Detailed polarization models  were constructed for each parameter-grid point, as set out in Section~\ref{sec:modover}, with results passband-integrated for comparison with the observations.    The resulting $\chi^2$ map is shown in Fig.~\ref{fig:JABchi2}.   Confidence intervals on this map were generated from bootstrapped datasets by evaluating best-fit inclinations for each \omomc\ from the $\chi^2$ 
maps, for 1000 samplings.  The 1-$\sigma$ ranges give the corresponding $\Delta\chi^2$, from which the confidence intervals may be inferred.

\begin{table*}
\caption{Stellar parameters for initial models (solid-body surface rotation), spanning the range of allowed inclinations.  For given $i$, the value of \omomc\ follows from eqtn.~\eqref{eq:valley} (excepting the `$\omega{2}\pm$' models), and thence \vesini\ from eqtn.~\eqref{eq:vom}.  The final three rows are the position angles of the stellar rotation axis and the interstellar polarization (each in the range 0:180$^\circ$), and the magnitude of peak polarization (eqtn.~\ref{eq:serk}). The `$\omega{2}+$' and `$\omega{2}-$' columns correspond to changes of $\pm2\sigma$ in $\omomc$ at the extremes of $i$.  The majority of the tabulated results were obtained by adopting the \hipparcos\  parallax, with the final two columns being for the \gaia\ parallax, at the extremes of allowed inclinations.} 
\centering
\tabcolsep 5 pt
\begin{tabular}{llrrrrrrrrrrrrrrrr}
	\toprule
 Parameter&Unit
 &\multicolumn{4}{c}{\quad\vhrulefill\;{\hipparcos}\;\vhrulefill\quad}&
 &\multicolumn{2}{c}{\quad\vhrulefill\;$\omega$2+\;\vhrulefill\quad}&
 &\multicolumn{2}{c}{\quad\vhrulefill\;$\omega$2$-$\;\vhrulefill\quad}&
 &\multicolumn{2}{c}{\quad\vhrulefill\;{\gaia}\;\vhrulefill\quad}\\
	\midrule
$i$ &$^\circ$  &  60 & 70 &  80 & 90      &&63&90&&58&90&&60&90\\
\omomc\ && 1.000 & 0.972 &  0.953 & 0.943 &&0.998&0.954&&0.999&0.931&&1.000&0.943 \\
\vesini\ &\kms &324 & 316 &  312 & 311    &&323&312&&324&309&&319&311\B\\ 
\hline
\teff\ &kK&9.02 & 9.44 & 9.63 & 9.68          &&9.14&9.71&&8.98&9.66&&9.02&9.68\T\\
\tpole\ &kK    &10.60 & 10.79 & 10.90 & 10.90 &&10.70&10.99&&10.53&10.82&&10.60&10.90 \\
\teq\ &kK & 6.47 & 8.19 & 8.56 & 8.68         &&7.05&8.62&&6.65&8.74&&6.45&8.69\B\\
\rpole&\rsun\ & 2.13 & 2.14 & 2.15 & 2.15     &&2.14&2.14&&2.12&2.16&&2.19&2.21\T\\
\req&\rsun\ & 3.14 & 2.84 & 2.76 & 2.73       &&3.09&2.76&&3.10&2.71&&3.23&2.81\\
\prot\ &hr& 10.20 & 10.23 & 10.58 & 10.68     &&10.33&10.73&&9.86&10.64&&10.66&10.97\\
$\thtbar$&mas & 0.97 & 0.91 & 0.89 & 0.89     &&0.96&0.89&&0.97&0.88&&0.97&0.89\B\\
\loggp &dex cgs& 4.17 & 4.19 & 4.18 & 4.18    &&4.16&4.17&&4.19&4.20&&4.14&4.17\T\\
\logge &dex cgs& 2.52 & 3.48 & 3.59 & 3.64    &&2.88&3.57&&2.71&3.70&&2.48&3.63\B\\
$M$&\msun& 2.42 & 2.58 & 2.55 & 2.57          &&2.40&2.46&&2.55&2.69&&2.41&2.64\T\\
\logL\ &dex & 1.63 & 1.67 & 1.69 & 1.70       &&1.65&1.70&&1.61&1.69&&1.65&1.72\B\\
\midrule
$\theta_*$&$^\circ$& 71.4 & 71.5 & 71.5 & 71.5&&71.4&71.5&&71.3&71.4\T \\
$\theta_{\rm i}$ & $^\circ$ & 56.8 & 55.4 & 53.7 & 53.7 &&58.1&54.1&&55.2&51.5\\
$p(\lambda_{\rm max})$ & ppm & 22.0 & 20.2 & 18.4 & 18.3 &&24.0&18.7&&20.2&16.6\\
\bottomrule
\end{tabular}
\label{tbl:params}
\end{table*}

\begin{table*}
\caption{Parameter sensitivity to fixed inputs, showing differences (model minus base) with respect to a reference model having
$i=80^\circ$, $\omomc=0.953$, $\vesini=312.3$~\kms, [M/H] = $-0.50$, \hipparcos\ parallax ($\pi = 39.28$~mas), $E(B-V) = 0\fm005$, and solid-body surface rotation.  }

\centering
\begin{tabular}{l l r@{.}l ccccccccccc}
\toprule
\multicolumn{2}{l}{Parameter}&\multicolumn{2}{c}{Base}   
                                     &    [M/H]   &$\pi$/mas   & $E(B-V)$   & \multicolumn{2}{c}{UV flux} 
                                                                                                      &\multicolumn{2}{c}{\vesini}
												                                &\multicolumn{2}{c}{\omomc} \\
&&\multicolumn{2}{c}{value}                             &   $=0.0$   &  $=38.23$  & $=0\fm0$     &$\times1.05$&   $/1.05$  & +5\kms\    & $-5$\kms\  &  $+0.01$  & $-0.01$ \\ 
\midrule
\teff             & kK      & 9&632  &  $+0.174$  &  $+0.002$  &  $-0.038$  &  $+0.091$  &  $-0.090$  &  $-0.001$  &  $+0.001$  &  $+0.020$  &  $-0.018$ \\
\loggp            & dex cgs & 4&180  &  $+0.014$  &  $-0.012$  &  $+0.001$  &  $+0.005$  &  $-0.005$  &  $+0.014$  &  $-0.014$  &  $-0.015$  &  $+0.015$ \\
$R_{\rm p}/\rsun$ &         & 2&148  &  $-0.067$  &  $+0.058$  &  $-0.005$  &  $-0.025$  &  $+0.025$  &  $+0.000$  &  $-0.001$  &  $-0.008$  &  $+0.007$ \\
$R_{\rm e}/\rsun$ &         & 2&762  &  $-0.087$  &  $+0.074$  &  $-0.007$  &  $-0.032$  &  $+0.032$  &  $+0.000$  &  $-0.001$  &  $+0.031$  &  $-0.028$ \\
$M/\msun$         &         & 2&547  &  $-0.079$  &  $+0.069$  &  $-0.006$  &  $-0.029$  &  $+0.030$  &  $+0.083$  &  $-0.081$  &  $-0.107$  &  $+0.104$ \\
\multicolumn{2}{l}{$\logL$} & 1&690  &  $+0.003$  &  $+0.023$  &  $-0.009$  &  $+0.006$  &  $-0.006$  &  $+0.000$  &  $+0.000$  &  $+0.007$  &  $-0.007$ \\
\prot\ & hr & 10&575 & $-0.330$ & +0.285 & $-0.025$ & $-0.122$ &  +0.125 & $-0.165$ & +0.170 & +0.055 & $-0.049$ \\
\bottomrule
\end{tabular}
\begin{flushleft}
\end{flushleft}
\label{tbl:sensit}
\end{table*}

As found in our previous studies of rapid rotators \citep{cotton17, bailey20b,lewis22}, the photo\-polarimetry is reproduced by a range of models, with \omomc\ decreasing with increasing axial inclination (thereby maintaining the overall effective image asymmetry required to generate the observed polarization signal).  
A simple analytical approximation to results of 
the best-fitting polarization models is given by
\begin{alignat}{4}
\label{eq:valley}
 &\omomc \simeq\;&& 0.961 - j(&&1.90\times10^{-3} -\, &&4.5\times10^{-5}j)
\end{alignat}
for $i \ge 60^\circ$, where $j = i - 75^\circ$ (dashed line in Fig.~\ref{fig:JABchi2}).   Corresponding physical parameters  for selected points along this 
locus are summarized in Table~\ref{tbl:params} for the initial parameter grid (columns 3--6).
There is insufficient asymmetry in the projected stellar image to generate the observed polarimetric signal for models having $i \lesssim 60^\circ$ or $\omomc \lesssim 0.93$.

\subsection{The rotation-period `problem'\ldots}
\label{sec:pprob}

Table~\ref{tbl:params} shows that, for the initial parameter grid, the surface rotation periods, \prot, along the $\chi^2$ `valley' of Fig.~\ref{fig:JABchi2} are in the range 10.2--10.7~hours (for $i = 60\rightarrow90^\circ$).  These values are
sufficiently close to the \textit{TESS} photo\-metric period ($\pfot =11.1$~hr; Section~\ref{sec:tess}) to suggest that the photo\-metric period may very well be the rotation period.

In part  to see if the model \prot\ values could be straightforwardly reconciled with \pfot, we conducted parameter-sensitivity tests, and also recalculated the parameter grid with the \gaia\ parallax (which, being smaller, leads to slightly larger radii, hence larger values of \prot\ for given \veq).   
Selected results are included in
Tables~\ref{tbl:params} 
(\gaia\ parallax, columns 11, 12)
and~\ref{tbl:sensit} (sensitivity tests).


The inferred effective temperature is, unsurprisingly, mildly sensitive (at the $\sim$100~K level) to [M/H], and to the value of the integrated UV flux, but other parameters (including \prot) appear to be quite robustly determined, with changes that are generally within the spread of values resulting from the $i$, \omomc\ indeterminancy.   This robustness propagates into the polarization modelling;  a solar-abundance test grid (with concomitant changes in all basic parameters) gives a $\chi^2$ map that is practically indistinguishable from that shown in Fig.~\ref{fig:JABchi2}, confirming our expectation that the modelled polarization is primarily sensitive to $i$ and to $\omomc$, with relatively little dependence on other parameters (within reasonable bounds).

Although confirming the obvious result that smaller parallax (greater distance, hence larger radius) and smaller \vesini\  should push \prot\ to larger values, the
model rotation periods in Tables~\ref{tbl:params} and~\ref{tbl:sensit} still all fall short of \pfot.   The differences are smallest for the highest-inclination, \gaia-parallax models, with the extreme $i=90^\circ$ model in Table~\ref{tbl:params} requiring a reduction of only $\sim$5~\kms\ in \vesini\ to force agreement.   However,
Fig.~\ref{fig:rotmont}(f) indicates that even as small a change as this is barely compatible with the line-profile analysis (for a continuum placement chosen to give consistency with solid-body surface rotation, which already results in lower \vesini\ values than would otherwise be found;  Fig.~\ref{fig:rotdiff}).

We conclude that the combined hypotheses of both (i)~solid-body surface rotation and (ii) $\pfot=\prot$, while not completely ruled out, are at best only marginally consistent with the data.

\subsection{{\ldots}and some possible resolutions}
\label{sec:psolve}

There are several possible circumstances that could address this modest discrepancy between \prot\ and \pfot.  For example, reducing the \gaia\ parallax by $\sim$2$\sigma$ would increase the distance, model radius, and hence \prot, by a further $\sim$2\%\ (although the parallax would then be $\sim$11$\sigma$ from the \hipparcos\ value).   Such `fine tuning' cannot be excluded;
nevertheless, we should also not discount the possibility that \pfot\ need not necessarily be identical to the solid-body rotation period.    

We note, for example, that both $g$- and $r$-mode pulsation periods can be in the same range as the rotation period.  
The attraction of pulsation as the origin of photometric variability is that does not require  time-variable magnetic fields to be invoked as a mechanism to generate rotational modulation through starspots (in a broad sense).   Where magnetic fields have been directly detected in OBA stars (which are thought to have primarily radiative envelopes), they appear to be fossil remnants, an interpretation consistent with associated highly reproducible, strictly periodic photometric and spectroscopic variability (e.g., \citealt{hubrig21}).
This behaviour contrasts with much of the low-amplitude variability revealed by space photometry and widely attributed to rotational modulation (e.g., \citealt{balona16}).

\citet{saio18} found a low-frequency `hump' in time-series power spectra 
at frequencies just below the rotation frequency, resulting from $r$-mode pulsation with azimuthal wavenumber \mbox{$|m| = 1$}.    Subsequently \citeauthor{lee20} (\citeyear{lee20};  \citealt{lee21, lee22}) have shown that
overstable convective modes in the core of an early-type star can couple with $g$~modes in the radiative envelope, provided the core rotates slightly faster than the envelope.  These modes are therefore candidates for the processes underlying the photometric variability.

However, the $r$-mode modelling suggests a more complex frequency spectrum than is exhibited by \zAql, while the periods driven by overstable convective modes should necessarily be shorter than the surface rotation periods (whereas we find to the contrary, that $\pfot \gtrsim \prot$).  
While pulsational variability close to the rotation frequency remains a possibility, arguments in favour of that hypothesis do not appear compelling at present.

\subsection{Differential rotation?}

If, instead, the photometric variability is attributed to rotational modulation of  surface-brightness inhomogeneities then differential surface rotation offers a straightforward solution to the discrepancy between the modelled (equatorial) rotation period, \prote, and \pfot:  spots could simply be at latitudes rotating with a different angular speed.

As described in Appendix~\ref{sec:lpm}, 
an ad hoc parametrization of differential rotation was considered as part of our initial examination of rotational velocities.   That analysis of the line-profile shape found no direct evidence for substantial differential rotation.
However, a purely empirical approach  cannot rule out lower-level differential rotation, and necessarily introduces an arbitrary element (i.e., an ad hoc characterization of the dependence of $\omega$ on colatitude $\theta$).

For a second phase of the analysis we therefore pursued a direct physical approach, using surface-rotation profiles, $\omega(\theta)$, generated from \mbox{\ester} 2-D stellar-structure models, as elaborated in Appendix~\ref{sec:estermod} (and illustrated in Fig.~\ref{fig:ester}).   

We first repeated the exercise of  establishing the relationship between \vesini\ and \omomc\
for these differentially-rotating models (cf.\ Appendices~\ref{sec:rotnumx},  \ref{sec:estermod}), then generated new parameter grids (for both \hipparcos\ and \gaia\ parallaxes).
Selected results are incorporated into Figs.~\ref{fig:pltGrid} and~\ref{fig:JABchi2}.
Fig.~\ref{fig:JABchi2} shows that  only the combination of \gaia\ parallax and \ester\ rotation profiles gives agreement between \prote\ and \pfot\ at $i$, \omomc\ values that are consistent with the polarimetry, without requiring any fine tuning.   

This result arises because \ester-type differential rotation has greatest angular velocity at temperate latitudes (Fig.~\ref{fig:ester}), leading to projected \textit{equatorial} rotation velocities that are generally $\sim1$--2\%\ smaller than rigid-rotator results
(Appendix~\ref{sec:estermod}).   
The relationships between \vesini\ and \omomc\ are essentially independent of parallax,
but with lower \veq\ values favoured by 
lower \omomc\ and by higher inclinations.
The combination of these effects, along with their non-linear dependences on \omomc, means that, within the $\chi^2$ valley of Fig.~\ref{fig:JABchi2}, \prote\ matches \pfot\ only for \gaia+\ester\ models, at high inclinations ($i \gtrsim 80^\circ$, such that  $\veq\simeq\vesini$) for the \omomc\ values indicated by the photo\-polarimetry.

Given the sensitivity of \prote\ to rather small changes in \vesini, this is clearly not a unique result (nor a strong validation) of the \ester\ rotation profile.  For example, an arbitrary surface-rotation law of the form of eqtn.~\ref{eq:adhoc}, with $\alpha \sim -0.03$, would give a similar outcome.   Nevertheless, the success of the (almost) `no free parameters' \ester\ models in bringing about agreement between \prote\ and \pfot\ is encouraging and suggestive. 

A further caveat is that Fig.~\ref{fig:JABchi2} is not fully self-consistent, as the various $\pfot = \prote$ lines are calculated under slightly different sets of assumptions, while the $\chi^2$ map is specifically for solid-body surface rotation and the \hipparcos\ parallax.   We have not addressed this minor inconsistency, for several reasons:
\begin{enumerate}
\item We consider the existing calculations already to be sufficient to demonstrate that the modelled \prote\ and observed \pfot\ can readily be reconciled;  while other stellar parameters are insensitive to details of the rotation profile and parallax.
\item We know the $\chi^2$ map is very insensitive to most input parameters (including small changes in parallax), excepting the $i$, $\omomc$ pair.  While any sensitivity to differential surface rotation is less clear, it is unlikely to be large, given the small departures from solid-body rotation implied by the \ester\ profiles (and the empirical limits on strongly differential rotation).
\item If the photo\-metric variability does arise from starspots, we don't know their latitude(s).  Although we have focussed on reconciling the \textit{equatorial} rotation period of the models with \pfot, the \textit{spot} rotation periods could easily be 1--2 per cent different (in either direction) in the presence of differential surface rotation.  As shown in Fig.~\ref{fig:JABchi2}, such small differences can have a relatively large effect on the location of the \prote\ locus  in the $i$, $\omomc$ plane, which is likely to dominate the uncertainties.
\end{enumerate}

\begin{table}
\caption{Stellar parameters for selected \gaia-parallax, \ester-rotation models
  (cp.\ Table~\ref{tbl:params}).   The listed models are indicated in Fig.~\ref{fig:JABchi2}, with the `base' model (adopted as our preferred solution) corresponding to the intersection of the $\chi^2$ valley  with the `$\prote=\pfot$' locus.}
\centering
\tabcolsep 5 pt
\begin{tabular}{llrrrrrrrrrrr}
	\toprule
 Parameter&Unit&\multicolumn{1}{c}{Base}&
\multicolumn{4}{c}{\quad\vhrulefill\;Variants\;\vhrulefill\quad}\\
	\midrule

$i$       & $^\circ$ &   84.9   &  82.4    &  77.9    &  90.0    &  90.0\T\\
\omomc\   &         &   0.946   &  0.931   &  0.968   &  0.957   &  0.925\\
\vesini\  & \kms    &   306     &  304     &  311     &  308     &  303\B\\
\hline
\teff\    & kK      &    9.693  &   9.638  &   9.645  &   9.741  &   9.661\T\\
\tpole\   & kK      &   10.952  &  10.817  &  11.020  &  11.063  &  10.814  \\
\teq      & kK      &    8.680  &   8.733  &   8.434  &   8.634  &   8.790\B \\
\rpole    & \rsun   &    2.21   &   2.22   &   2.19   &   2.20   &   2.22\T \\
\req      & \rsun   &    2.82   &   2.78   &   2.89   &   2.84   &   2.77   \\
\prote\    & hr      &   11.12   &  11.00   &  11.00   &  11.20   &  11.07   \\
$\thtbar$ & mas     &    0.89   &   0.88   &   0.90   &   0.89   &   0.88\B \\
\loggp    & dex cgs & 4.15      &4.18      &4.14      &4.14      &4.18\T \\
\logge    & dex cgs &    3.60   &   3.68   &   3.47   &   3.53   &   3.70\B \\
$M$       & \msun   &    2.53   &   2.71   &   2.42   &   2.41   &   2.73\T \\
\logL     & dex     &    1.72   &   1.71   &   1.73   &   1.74   &   1.71\B \\
\bottomrule
\end{tabular}
\label{tbl:params2}
\end{table}

\begin{figure*}
\includegraphics[width=\textwidth]{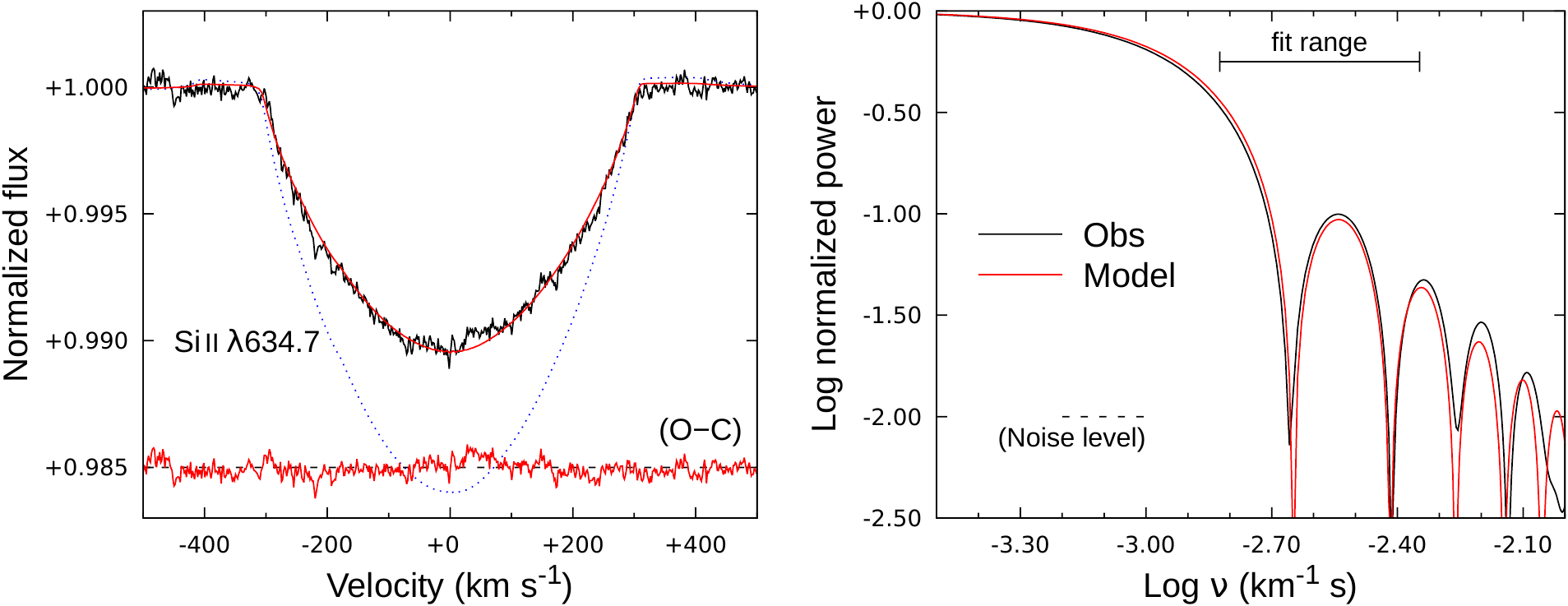}
\caption{Left: observed and modelled Si\;{\sc ii} profiles
for the base model of Table~\ref{tbl:params2} ($\vesini = 306$~\kms; \mbox{\ester} rotation, \mbox{\gaia} parallax).
The model line depth has been scaled by 0.95$\times$ to facilitate comparison of line shapes; the dotted blue line shows an otherwise identical model (including the ad hoc scaling) for solar abundances.  Right:  the normalized Fourier transforms, showing the frequency range over which the observed and modelled transforms were compared (Section~\ref{sec:lpm}); the white-noise power level is indicated. }
\label{fig:rotprof}
\end{figure*}

\begin{figure}
\includegraphics[width=\columnwidth]{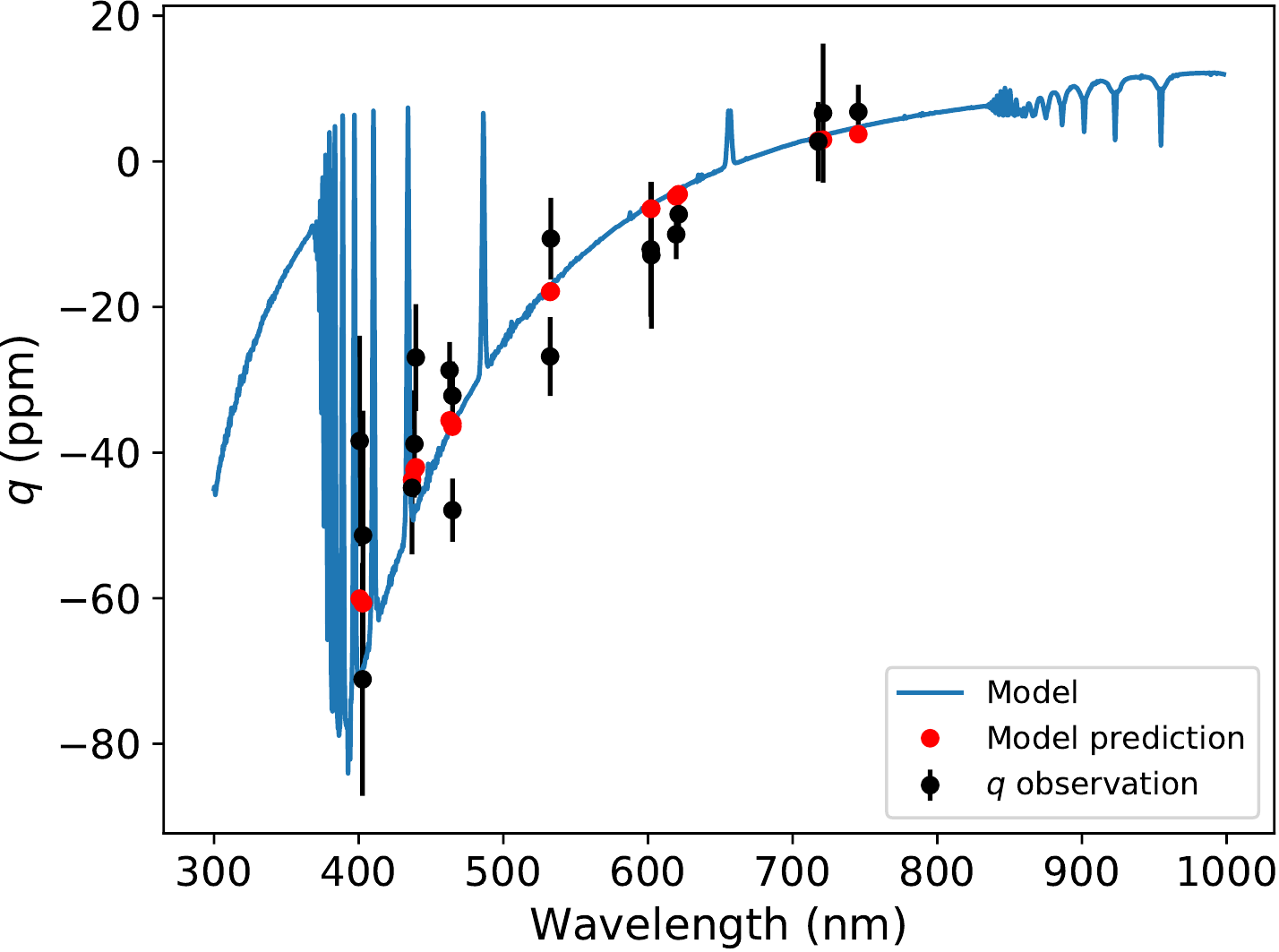}
\caption{Comparison of observed and modelled polarizations.   Red dots are passband-integrated model results.   The `observed' values have been corrected for foreground interstellar polarization, and rotated so that the polarization is entirely in $q$.  The model is for solid-body surface rotation at $i=85^\circ$, $\omomc=0.95$.  }
\label{fig:polplt}
\end{figure}

\begin{figure}
\includegraphics[width=\columnwidth]{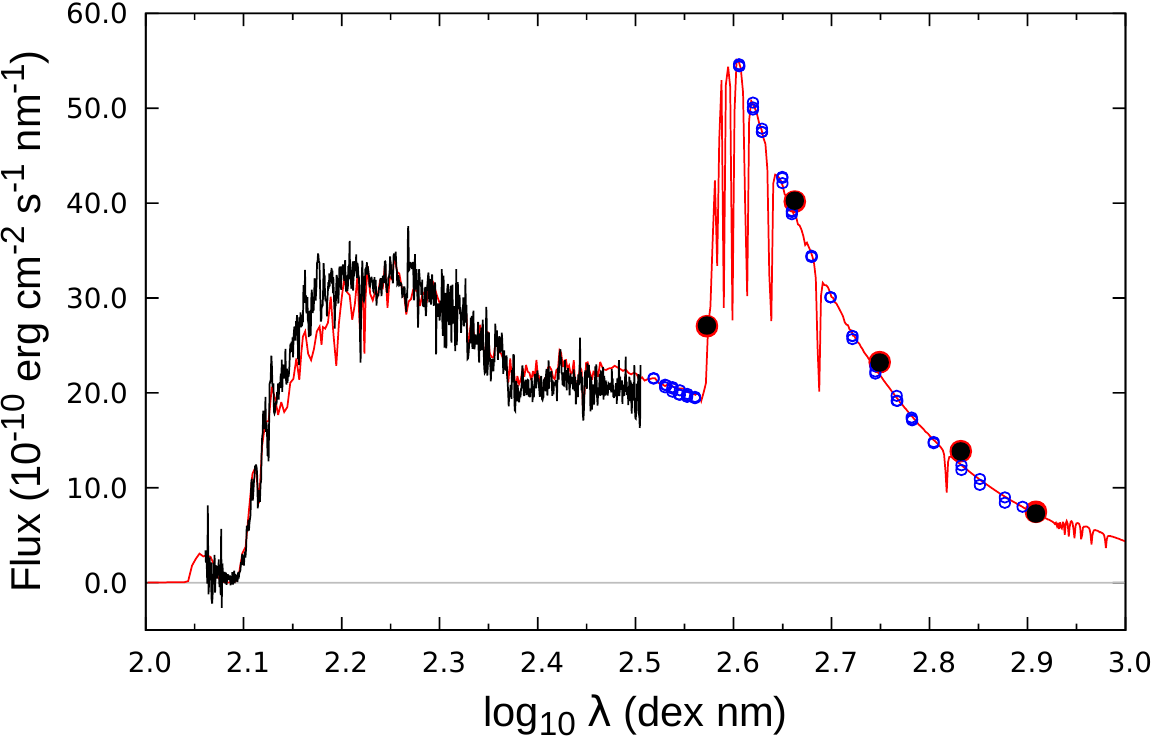}
\caption{Comparison of measured and modelled fluxes.   Observed \textit{IUE} and \textit{UBVRI} fluxes are shown in black.   Optical spectro\-photometry from  \citet{breger76} and \citet{adelman80}, normalized at 500nm, is shown as small blue dots (but was not used in the modelling).
The `base' model of Table~\ref{tbl:params2}, reddened with $E(B-V) = 0\fm005$, is shown in red.}  
\label{fig:flxplt}
\end{figure}

\section{Discussion}
\label{sec:disco}

Selected numerical results for the \gaia+\ester\ parameter grid are given in Table~\ref{tbl:params2}
(other parameter grids give results intermediate  
between those in Tables~\ref{tbl:params} and~\ref{tbl:params2}).   We take 
the `base' model listed there, for which $\prote=\pfot$, as our adopted specific solution.  
If starspots do give rise to the photometric variability, then the combination of high axial inclination and $\sim$continuous variation implies that they must have an extensive distribution in longitude.

\begin{figure*}
\centering
\includegraphics[width=\columnwidth]{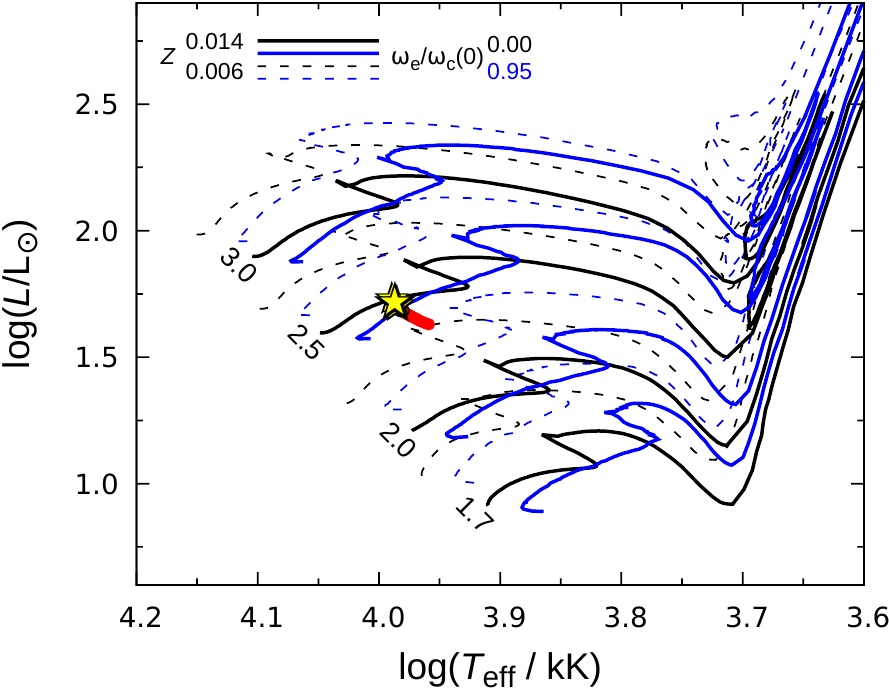}
\includegraphics[width=\columnwidth]{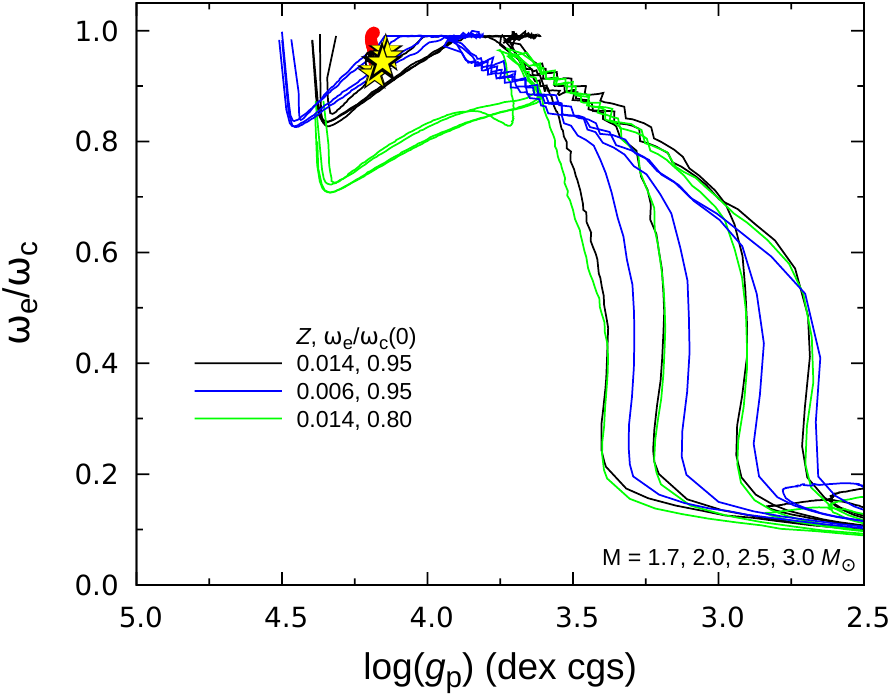}
\caption{Left: Hertzsprung-Russell diagram.   Evolutionary tracks are from \citet{georgy13} for the indicated ZAMS masses (in solar units), metallicities $Z$, and initial (ZAMS) \omomc\ values.   
Yellow stars show \textit{all} the solutions from Table~\ref{tbl:params2}, although they are almost inseparable in this plot.
Red dots show results from Table~\ref{tbl:params}; additional solutions from the sensitivity tests
(Table~\ref{tbl:sensit}) all fall under the yellow stars.   Right: evolution of model \omomc\ values.   Overall, evolution is from higher to lower gravities;  the $\sim$horizontal regions at $\loggp \simeq 4$ correspond to the main-sequence phase. }
\label{fig:hrd}
\end{figure*}

The line-profile modelling for the adopted solution is shown in Fig.~\ref{fig:rotprof},
the predicted and observed polar\-izations in Fig.~\ref{fig:polplt}, and the flux distributions in Fig.~\ref{fig:flxplt}.   All these comparisons show satisfactory agreement between models and observations.

The mean angular diameters predicted by  the tabulated models are in excellent accord with the interferometric value given by 
\citet[$0.895 \pm 0.017$~mas]{boyajian12}, but are inconsistent with the  interim analysis reported by \citeauthor{peterson06}
(\citeyear{peterson06};  cf.~Table~\ref{tbl:basic});  their values of $i = 90^\circ$, $\omomc = 0.99$, $\theta_* = 45^\circ$ are also at odds with the photo\-polarimetric results (Table~\ref{tbl:params}).\footnote{The \mbox{\tpole, \omomc, $i$} triplet reported by \mbox{\citeauthor{peterson06}} requires  $\theta_{\rm P} \simeq 0.75$~mas ($\mbox{\thtbar} \simeq 0.88$~mas) to reproduce the observed $V$ magnitude.  The disagreement with their published value,
$\theta_{\rm P} = 0.815 \pm 0.005$~mas,
suggests that there may be typographical errors in their tabulated numbers.}


\subsection{Comparison with evolutionary models}

We compare our empirical results with models of the evolution of rotating stars from \citet{georgy13} in Fig.~\ref{fig:hrd}.\footnote{It is an early version of this comparison that underpinned the choice of parameters adopted for the \ester\ modelling described in Section~\ref{sec:estermod}.}    Their grids are for a range of ZAMS rotation rates, \omomc(0), and include metallicities representative of solar and LMC abundances ($Z = 0.006$, corresponding to $\mbox{[M/H]} \simeq -0.4$).   All our empirically determined masses fall within the range 
$M=2.53^{+0.20}_{-0.13}\msun$, in excellent agreement with the evolutionary mass for the solar-abundance tracks.   The LMC-abundance tracks suggest evolutionary masses $\sim$0.3\msun\ lower, barely consistent with empirical values.

Main-sequence A-type stars show a range of surface-abundance anomalies, usually involving selective  metal enhancements (the Am, Ap, and HgMn stars); only stars in the $\lambda$~Boo class are noted for their metal depletions.  This group is also characterized by relatively rapid rotation.  While \zAqlx\ is not a classic $\lambda$~Boo star in terms of its spectral morphology (Gray, personal communication), its subsolar metallicity may arise through a similar mechanism, generally thought to involve photo\-spheric accretion of depleted gas (e.g., \citealt{venn90}, \citealt{jermyn18}).   In that case, we would expect solar abundances to be more relevant to its evolution, as found in these comparisons

At these masses the evolutionary tracks are not strongly sensitive to the precise value of \omomc(0), although a high value is, of course, required for \zAql.  A ZAMS value close to $\sim$0.95 is  consistent with observations (Fig.~\ref{fig:hrd}, right-hand panel).

\section{Summary and conclusions}

We have presented new, very precise photo\-polarimetry of \zAqlx\ (Table~\ref{tbl:obs}).   Modelling those observations, together with supplementary analyses of the flux distribution and rotational velocity, allows us to determine the locus of allowed combinations of $i$ and \omomc\
(Fig.~\ref{fig:JABchi2}).   The polarimetry alone cannot break the degeneracy between these two parameters, but limits their values to
$i \gtrsim 60^\circ$, $\omomc \gtrsim 0.93$.

Periodic photo\-metric variability, demonstrated here for the first time (from \textit{TESS} observations), provides additional constraints under the plausible assumption that the newly established photo\-metric 
period, $\pfot = 11.12$~hr, can be identified with the rotation period.   
The rotation periods of
models based on rigid-body surface rotation are only marginally consistent with \pfot, requiring extreme values and fine tuning of para\-meters to push \ve\ and/or the parallax to appropriately low values.
However, model equatorial rotation periods are found to be in good agreement with \pfot\ for the combination of \gaia\ parallax and the differential surface rotation predicted by \ester\ models.

The inferred physical parameters of \zAql\ are quite insensitive to these issues, as demonstrated by the small range of solutions listed in Tables~\ref{tbl:params}--\ref{tbl:params2};  our adopted specific characterization is given in column~3 of Table~\ref{tbl:params2}.   Taking the full ranges of parameter values in that Table as a reasonably conservative estimate of the 1-$\sigma$ uncertainties, we find
\mbox{$M = 2.53\pm0.16\,\msun$}, \mbox{$\logL = 1.72\pm0.02$,}
\mbox{$\rpole = 2.21\pm 0.02\,\rsun$}, \mbox{$\teff = 9693 \pm 50$~K}, \mbox{$i = 85{^{+5}_{-7}}^\circ$}, and \mbox{$\omomc = 0.95\pm0.02$}.

Comparison of our results with grids of single-star evolution calculations shows excellent agreement for solar-abundance models, but poorer agreement with models at lower metallicities that approximately match the depleted surface abundances.   This suggests that the observed photo\-spheric depletions may not be global, but instead confined only to the surface layers.

\section*{Acknowledgements}

This paper is based in large part on data obtained with the Anglo-Australian Telescope at Siding Spring Observatory; we acknowledge the traditional owners of the land on which the AAT stands, the Gamilaraay people, and we pay our respects to elders past and present.
We made use of the Washington Double Star Catalog, maintained at the U.S.\ Naval Observatory, as well as observations made with the \textit{International Ultraviolet Explorer} and \textit{TESS} satellites, obtained from the MAST data archive at the Space Telescope Science Institute, which is operated by the Association of Universities for Research in Astronomy, Inc., under NASA contract NAS 5–26555.  Funding for the \textit{TESS} mission is provided by the NASA Explorer Program. 
CFHT data were accessed by using the facilities of the Canadian Astronomy Data Centre, operated by the National Research Council of Canada with the support of the Canadian Space Agency.
We also benefitted from NASA's Astrophysics Data System bibliographic service,
and the SIMBAD database, operated at CDS, Strasbourg, France.  
We thank 
Nicholas Borsato,  Dag Evensberget, Behrooz Karamiqucham, Jonathan Marshall, and Jinglin Zhao for  contributions to observing runs, our anonymous referee for useful remarks, and Conny Aerts, Derek Buzasi, Richard Gray, and Michel Rieutord for helpful correspondence.
DVC thanks the Friends of MIRA for their support.

\section*{Data Availability}

The new polarization data used for this project are listed in Table~\ref{tbl:obs}.   All other data are from publicly accessible archives.



\bibliographystyle{mnras}
\bibliography{zetaAql}

\newcommand{\noop}[1]{}
\begin{thebibliography}{}
\makeatletter
\relax
\def\mn@urlcharsother{\let\do\@makeother \do\$\do\&\do\#\do\^\do\_\do\%\do\~}
\def\mn@doi{\begingroup\mn@urlcharsother \@ifnextchar [ {\mn@doi@}
  {\mn@doi@[]}}
\def\mn@doi@[#1]#2{\def\@tempa{#1}\ifx\@tempa\@empty \href
  {http://dx.doi.org/#2} {doi:#2}\else \href {http://dx.doi.org/#2} {#1}\fi
  \endgroup}
\def\mn@eprint#1#2{\mn@eprint@#1:#2::\@nil}
\def\mn@eprint@arXiv#1{\href {http://arxiv.org/abs/#1} {{\tt arXiv:#1}}}
\def\mn@eprint@dblp#1{\href {http://dblp.uni-trier.de/rec/bibtex/#1.xml}
  {dblp:#1}}
\def\mn@eprint@#1:#2:#3:#4\@nil{\def\@tempa {#1}\def\@tempb {#2}\def\@tempc
  {#3}\ifx \@tempc \@empty \let \@tempc \@tempb \let \@tempb \@tempa \fi \ifx
  \@tempb \@empty \def\@tempb {arXiv}\fi \@ifundefined
  {mn@eprint@\@tempb}{\@tempb:\@tempc}{\expandafter \expandafter \csname
  mn@eprint@\@tempb\endcsname \expandafter{\@tempc}}}

\bibitem[\protect\citeauthoryear{{Absil} et~al.,}{{Absil}
  et~al.}{2008}]{absil08}
{Absil} O.,  et~al., 2008, \mn@doi [\aap] {10.1051/0004-6361:200810008}, \href
  {https://ui.adsabs.harvard.edu/abs/2008A&A...487.1041A} {487, 1041}

\bibitem[\protect\citeauthoryear{{Abt} \& {Morrell}}{{Abt} \&
  {Morrell}}{1995}]{abt95}
{Abt} H.~A.,  {Morrell} N.~I.,  1995, \mn@doi [\apjs] {10.1086/192182}, \href
  {https://ui.adsabs.harvard.edu/abs/1995ApJS...99..135A} {99, 135}

\bibitem[\protect\citeauthoryear{{Adelman}, {Pyper}  \& {White}}{{Adelman}
  et~al.}{1980}]{adelman80}
{Adelman} S.~J.,  {Pyper} D.~M.,   {White} R.~E.,  1980, \mn@doi [\apjs]
  {10.1086/190679}, \href
  {https://ui.adsabs.harvard.edu/abs/1980ApJS...43..491A} {43, 491}

\bibitem[\protect\citeauthoryear{{Ammler-von Eiff} \& {Reiners}}{{Ammler-von
  Eiff} \& {Reiners}}{2012}]{ammler12}
{Ammler-von Eiff} M.,  {Reiners} A.,  2012, \mn@doi [\aap]
  {10.1051/0004-6361/201118724}, \href
  {https://ui.adsabs.harvard.edu/abs/2012A&A...542A.116A} {542, A116}

\bibitem[\protect\citeauthoryear{{Bailer-Jones}, {Rybizki}, {Fouesneau},
  {Demleitner}  \& {Andrae}}{{Bailer-Jones} et~al.}{2021}]{bailerjones21}
{Bailer-Jones} C.~A.~L.,  {Rybizki} J.,  {Fouesneau} M.,  {Demleitner} M.,
  {Andrae} R.,  2021, \mn@doi [\aj] {10.3847/1538-3881/abd806}, \href
  {https://ui.adsabs.harvard.edu/abs/2021AJ....161..147B} {161, 147}

\bibitem[\protect\citeauthoryear{{Bailey}, {Lucas}  \& {Hough}}{{Bailey}
  et~al.}{2010}]{bailey10}
{Bailey} J.,  {Lucas} P.~W.,   {Hough} J.~H.,  2010, \mn@doi [\mnras]
  {10.1111/j.1365-2966.2010.16634.x}, \href
  {https://ui.adsabs.harvard.edu/abs/2010MNRAS.405.2570B} {405, 2570}

\bibitem[\protect\citeauthoryear{{Bailey}, {Kedziora-Chudczer}, {Cotton},
  {Bott}, {Hough}  \& {Lucas}}{{Bailey} et~al.}{2015}]{bailey15}
{Bailey} J.,  {Kedziora-Chudczer} L.,  {Cotton} D.~V.,  {Bott} K.,  {Hough}
  J.~H.,   {Lucas} P.~W.,  2015, \mn@doi [\mnras] {10.1093/mnras/stv519}, \href
  {https://ui.adsabs.harvard.edu/abs/2015MNRAS.449.3064B} {449, 3064}

\bibitem[\protect\citeauthoryear{{Bailey}, {Cotton}, {Kedziora-Chudczer}, {De
  Horta}  \& {Maybour}}{{Bailey} et~al.}{2020a}]{bailey20a}
{Bailey} J.,  {Cotton} D.~V.,  {Kedziora-Chudczer} L.,  {De Horta} A.,
  {Maybour} D.,  2020a, \mn@doi [\pasa] {10.1017/pasa.2019.45}, \href
  {https://ui.adsabs.harvard.edu/abs/2020PASA...37....4B} {37, e004}

\bibitem[\protect\citeauthoryear{{Bailey}, {Cotton}, {Howarth}, {Lewis}  \&
  {Kedziora-Chudczer}}{{Bailey} et~al.}{2020b}]{bailey20b}
{Bailey} J.,  {Cotton} D.~V.,  {Howarth} I.~D.,  {Lewis} F.,
  {Kedziora-Chudczer} L.,  2020b, \mn@doi [\mnras] {10.1093/mnras/staa785},
  \href {https://ui.adsabs.harvard.edu/abs/2020MNRAS.494.2254B} {494, 2254}

\bibitem[\protect\citeauthoryear{{Bailey} et~al.,}{{Bailey}
  et~al.}{2021}]{bailey21}
{Bailey} J.,  et~al., 2021, \mn@doi [\mnras] {10.1093/mnras/stab172}, \href
  {https://ui.adsabs.harvard.edu/abs/2021MNRAS.502.2331B} {502, 2331}

\bibitem[\protect\citeauthoryear{{Balona} \& {Abedigamba}}{{Balona} \&
  {Abedigamba}}{2016}]{balona16}
{Balona} L.~A.,  {Abedigamba} O.~P.,  2016, \mn@doi [\mnras]
  {10.1093/mnras/stw1443}, \href
  {https://ui.adsabs.harvard.edu/abs/2016MNRAS.461..497B} {461, 497}

\bibitem[\protect\citeauthoryear{{Blackwell} \& {Shallis}}{{Blackwell} \&
  {Shallis}}{1977}]{blackwell77}
{Blackwell} D.~E.,  {Shallis} M.~J.,  1977, \mn@doi [\mnras]
  {10.1093/mnras/180.2.177}, \href
  {https://ui.adsabs.harvard.edu/abs/1977MNRAS.180..177B} {180, 177}

\bibitem[\protect\citeauthoryear{{Boss}}{{Boss}}{1910}]{boss10}
{Boss} L.,  1910, Preliminary General Catalogue.
Carnegie Institution, Washington, DC

\bibitem[\protect\citeauthoryear{{Boss}}{{Boss}}{1937}]{boss37}
{Boss} B.,  1937, General Catalogue of 33342 stars for the epoch 1950.
Carnegie Institution, Washington, DC

\bibitem[\protect\citeauthoryear{{Boyajian} et~al.,}{{Boyajian}
  et~al.}{2012}]{boyajian12}
{Boyajian} T.~S.,  et~al., 2012, \mn@doi [\apj] {10.1088/0004-637X/746/1/101},
  \href {https://ui.adsabs.harvard.edu/abs/2012ApJ...746..101B} {746, 101}

\bibitem[\protect\citeauthoryear{{Boyarchuk} \& {Kopylov}}{{Boyarchuk} \&
  {Kopylov}}{1964}]{boyarchuk64}
{Boyarchuk} M.~E.,  {Kopylov} I.,  1964, \izkry, 31, 44

\bibitem[\protect\citeauthoryear{{Breger}}{{Breger}}{1976}]{breger76}
{Breger} M.,  1976, \mn@doi [\apjs] {10.1086/190392}, \href
  {https://ui.adsabs.harvard.edu/abs/1976ApJS...32....7B} {32, 7}

\bibitem[\protect\citeauthoryear{{Burnham}}{{Burnham}}{1874}]{Burnham74}
{Burnham} S.~W.,  1874, \mn@doi [\mnras] {10.1093/mnras/35.1.31}, \href
  {https://ui.adsabs.harvard.edu/abs/1874MNRAS..35...31B} {35, 31}

\bibitem[\protect\citeauthoryear{Carrington}{Carrington}{1863}]{carrington63}
Carrington R.~C.,  1863, Observations of the spots on the Sun: from November 9,
  1853, to March 24, 1861, made at Redhill.
Williams and Norgate

\bibitem[\protect\citeauthoryear{{Carroll}}{{Carroll}}{1933}]{Carroll33}
{Carroll} J.~A.,  1933, \mn@doi [\mnras] {10.1093/mnras/93.7.478}, \href
  {https://ui.adsabs.harvard.edu/abs/1933MNRAS..93..478C} {93, 478}

\bibitem[\protect\citeauthoryear{{Castelli} \& {Kurucz}}{{Castelli} \&
  {Kurucz}}{2003}]{castelli03}
{Castelli} F.,  {Kurucz} R.~L.,  2003, in {Piskunov} N.,  {Weiss} W.~W.,
  {Gray} D.~F.,  eds,  IAU Symposium Vol. 210, Modelling of Stellar
  Atmospheres. p.~A20 (\mn@eprint {arXiv} {astro-ph/0405087})

\bibitem[\protect\citeauthoryear{{Cochetti}, {Zorec}, {Cidale}, {Arias},
  {Aidelman}, {Torres}, {Fr{\'e}mat}  \& {Granada}}{{Cochetti}
  et~al.}{2020}]{cochetti20}
{Cochetti} Y.~R.,  {Zorec} J.,  {Cidale} L.~S.,  {Arias} M.~L.,  {Aidelman} Y.,
   {Torres} A.~F.,  {Fr{\'e}mat} Y.,   {Granada} A.,  2020, \mn@doi [\aap]
  {10.1051/0004-6361/201936444}, \href
  {https://ui.adsabs.harvard.edu/abs/2020A&A...634A..18C} {634, A18}

\bibitem[\protect\citeauthoryear{{Collins}}{{Collins}}{1963}]{collins63}
{Collins} George~W. I.,  1963, \mn@doi [\apj] {10.1086/147712}, \href
  {https://ui.adsabs.harvard.edu/abs/1963ApJ...138.1134C} {138, 1134; erratum
  139, 1401}

\bibitem[\protect\citeauthoryear{{Cotton}, {Bailey}, {Howarth}, {Bott},
  {Kedziora-Chudczer}, {Lucas}  \& {Hough}}{{Cotton} et~al.}{2017a}]{cotton17}
{Cotton} D.~V.,  {Bailey} J.,  {Howarth} I.~D.,  {Bott} K.,
  {Kedziora-Chudczer} L.,  {Lucas} P.~W.,   {Hough} J.~H.,  2017a, \mn@doi
  [Nature Astronomy] {10.1038/s41550-017-0238-6}, \href
  {http://adsabs.harvard.edu/abs/2017NatAs...1..690C} {1, 690}

\bibitem[\protect\citeauthoryear{{Cotton}, {Marshall}, {Bailey},
  {Kedziora-Chudczer}, {Bott}, {Marsden}  \& {Carter}}{{Cotton}
  et~al.}{2017b}]{cotton17b}
{Cotton} D.~V.,  {Marshall} J.~P.,  {Bailey} J.,  {Kedziora-Chudczer} L.,
  {Bott} K.,  {Marsden} S.~C.,   {Carter} B.~D.,  2017b, \mn@doi [\mnras]
  {10.1093/mnras/stx068}, \href
  {http://cdsads.u-strasbg.fr/abs/2017MNRAS.467..873C} {467, 873}

\bibitem[\protect\citeauthoryear{{Cotton} et~al.,}{{Cotton}
  et~al.}{2019}]{cotton19b}
{Cotton} D.~V.,  et~al., 2019, \mn@doi [\mnras] {10.1093/mnras/sty3318}, \href
  {http://adsabs.harvard.edu/abs/2019MNRAS.483.3636C} {483, 3636}

\bibitem[\protect\citeauthoryear{{De Rosa} et~al.,}{{De Rosa}
  et~al.}{2014}]{deRosa14}
{De Rosa} R.~J.,  et~al., 2014, \mn@doi [\mnras] {10.1093/mnras/stt1932}, \href
  {https://ui.adsabs.harvard.edu/abs/2014MNRAS.437.1216D} {437, 1216}

\bibitem[\protect\citeauthoryear{{Dravins}, {Lindegren}  \&
  {Torkelsson}}{{Dravins} et~al.}{1990}]{dravins90}
{Dravins} D.,  {Lindegren} L.,   {Torkelsson} U.,  1990, \aap, \href
  {https://ui.adsabs.harvard.edu/abs/1990A&A...237..137D} {237, 137}

\bibitem[\protect\citeauthoryear{{Espinosa Lara} \& {Rieutord}}{{Espinosa Lara}
  \& {Rieutord}}{2011}]{espinosa11}
{Espinosa Lara} F.,  {Rieutord} M.,  2011, \mn@doi [\aap]
  {10.1051/0004-6361/201117252}, \href
  {https://ui.adsabs.harvard.edu/abs/2011A&A...533A..43E} {533, A43}

\bibitem[\protect\citeauthoryear{{Espinosa Lara} \& {Rieutord}}{{Espinosa Lara}
  \& {Rieutord}}{2013}]{espinosa13}
{Espinosa Lara} F.,  {Rieutord} M.,  2013, \mn@doi [\aap]
  {10.1051/0004-6361/201220844}, \href
  {https://ui.adsabs.harvard.edu/abs/2013A&A...552A..35E} {552, A35}

\bibitem[\protect\citeauthoryear{{Fabricius} et~al.,}{{Fabricius}
  et~al.}{2021}]{fabricius21}
{Fabricius} C.,  et~al., 2021, \mn@doi [\aap] {10.1051/0004-6361/202039834},
  \href {https://ui.adsabs.harvard.edu/abs/2021A&A...649A...5F} {649, A5}

\bibitem[\protect\citeauthoryear{{Ferraz-Mello}}{{Ferraz-Mello}}{1981}]{ferraz81}
{Ferraz-Mello} S.,  1981, \mn@doi [\aj] {10.1086/112924}, \href
  {https://ui.adsabs.harvard.edu/abs/1981AJ.....86..619F} {86, 619}

\bibitem[\protect\citeauthoryear{{Gaia Collaboration}}{{Gaia
  Collaboration}}{2021a}]{gaia3a}
{Gaia Collaboration} 2021a, \mn@doi [\aap] {10.1051/0004-6361/202039657}, \href
  {https://ui.adsabs.harvard.edu/abs/2021A&A...649A...1G} {649, A1}

\bibitem[\protect\citeauthoryear{{Gaia Collaboration}}{{Gaia
  Collaboration}}{2021b}]{gaia3b}
{Gaia Collaboration} 2021b, \mn@doi [\aap] {10.1051/0004-6361/202039657e},
  \href {https://ui.adsabs.harvard.edu/abs/2021A&A...650C...3G} {650, C3}

\bibitem[\protect\citeauthoryear{{Georgy}, {Ekstr{\"o}m}, {Granada}, {Meynet},
  {Mowlavi}, {Eggenberger}  \& {Maeder}}{{Georgy} et~al.}{2013}]{georgy13}
{Georgy} C.,  {Ekstr{\"o}m} S.,  {Granada} A.,  {Meynet} G.,  {Mowlavi} N.,
  {Eggenberger} P.,   {Maeder} A.,  2013, \mn@doi [\aap]
  {10.1051/0004-6361/201220558}, \href
  {https://ui.adsabs.harvard.edu/abs/2013A&A...553A..24G} {553, A24}

\bibitem[\protect\citeauthoryear{{Gray}, {Corbally}, {Garrison}, {McFadden}  \&
  {Robinson}}{{Gray} et~al.}{2003}]{gray03}
{Gray} R.~O.,  {Corbally} C.~J.,  {Garrison} R.~F.,  {McFadden} M.~T.,
  {Robinson} P.~E.,  2003, \mn@doi [\aj] {10.1086/378365}, \href
  {https://ui.adsabs.harvard.edu/abs/2003AJ....126.2048G} {126, 2048}

\bibitem[\protect\citeauthoryear{{H{\"a}ggkvist} \& {Oja}}{{H{\"a}ggkvist} \&
  {Oja}}{1969}]{hoggkvist69}
{H{\"a}ggkvist} L.,  {Oja} T.,  1969, Arkiv for Astronomi, \href
  {https://ui.adsabs.harvard.edu/abs/1969ArA.....5..303H} {5, 303}

\bibitem[\protect\citeauthoryear{{Hill}}{{Hill}}{1982}]{hill82}
{Hill} G.,  1982, Publications of the Dominion Astrophysical Observatory
  Victoria, \href {https://ui.adsabs.harvard.edu/abs/1982PDAO...16...67H} {16,
  67}

\bibitem[\protect\citeauthoryear{{Hough}, {Lucas}, {Bailey}, {Tamura}, {Hirst},
  {Harrison}  \& {Bartholomew-Biggs}}{{Hough} et~al.}{2006}]{hough06}
{Hough} J.~H.,  {Lucas} P.~W.,  {Bailey} J.~A.,  {Tamura} M.,  {Hirst} E.,
  {Harrison} D.,   {Bartholomew-Biggs} M.,  2006, \mn@doi [\pasp]
  {10.1086/507955}, \href
  {https://ui.adsabs.harvard.edu/abs/2006PASP..118.1302H} {118, 1302}

\bibitem[\protect\citeauthoryear{{Howarth}}{{Howarth}}{2011}]{howarth11}
{Howarth} I.~D.,  2011, \mn@doi [\mnras] {10.1111/j.1365-2966.2011.18122.x},
  \href {https://ui.adsabs.harvard.edu/abs/2011MNRAS.413.1515H} {413, 1515}

\bibitem[\protect\citeauthoryear{{Howarth}}{{Howarth}}{2016}]{howarth16}
{Howarth} I.~D.,  2016, \mn@doi [\mnras] {10.1093/mnras/stw245}, \href
  {https://ui.adsabs.harvard.edu/abs/2016MNRAS.457.3769H} {457, 3769}

\bibitem[\protect\citeauthoryear{{Hubeny}}{{Hubeny}}{2012}]{hubeny12}
{Hubeny} I.,  2012, in {Richards} M.~T.,  {Hubeny} I.,  eds,  IAU Symposium
  Vol. 282, From Interacting Binaries to Exoplanets: Essential Modeling Tools.
  Cambridge University Press, pp 221--228

\bibitem[\protect\citeauthoryear{{Hubeny}, {Stefl}  \& {Harmanec}}{{Hubeny}
  et~al.}{1985}]{hubeny85}
{Hubeny} I.,  {Stefl} S.,   {Harmanec} P.,  1985, Bulletin of the Astronomical
  Institutes of Czechoslovakia, \href
  {https://ui.adsabs.harvard.edu/abs/1985BAICz..36..214H} {36, 214}

\bibitem[\protect\citeauthoryear{{Hubrig} \& {Sch{\"o}ller}}{{Hubrig} \&
  {Sch{\"o}ller}}{2021}]{hubrig21}
{Hubrig} S.,  {Sch{\"o}ller} M.,  2021, {Magnetic Fields in O, B, and A Stars}.
IoP Publishing

\bibitem[\protect\citeauthoryear{{Jermyn} \& {Kama}}{{Jermyn} \&
  {Kama}}{2018}]{jermyn18}
{Jermyn} A.~S.,  {Kama} M.,  2018, \mn@doi [\mnras] {10.1093/mnras/sty429},
  \href {https://ui.adsabs.harvard.edu/abs/2018MNRAS.476.4418J} {476, 4418}

\bibitem[\protect\citeauthoryear{{Johnson}, {Mitchell}, {Iriarte}  \&
  {Wisniewski}}{{Johnson} et~al.}{1966}]{johnson66}
{Johnson} H.~L.,  {Mitchell} R.~I.,  {Iriarte} B.,   {Wisniewski} W.~Z.,  1966,
  Communications of the Lunar and Planetary Laboratory, \href
  {https://ui.adsabs.harvard.edu/abs/1966CoLPL...4...99J} {4, 99}

\bibitem[\protect\citeauthoryear{{Kawaler}}{{Kawaler}}{2021}]{kawaler21}
{Kawaler} S.~D.,  2021, \mn@doi [Research Notes of the American Astronomical
  Society] {10.3847/2515-5172/ac351a}, \href
  {https://ui.adsabs.harvard.edu/abs/2021RNAAS...5..258K} {5, 258}

\bibitem[\protect\citeauthoryear{{Lee}}{{Lee}}{2021}]{lee21}
{Lee} U.,  2021, \mn@doi [\mnras] {10.1093/mnras/stab1433}, \href
  {https://ui.adsabs.harvard.edu/abs/2021MNRAS.505.1495L} {505, 1495}

\bibitem[\protect\citeauthoryear{{Lee}}{{Lee}}{2022}]{lee22}
{Lee} U.,  2022, \mn@doi [\mnras] {10.1093/mnras/stac1021}, \href
  {https://ui.adsabs.harvard.edu/abs/2022MNRAS.513.2522L} {513, 2522}

\bibitem[\protect\citeauthoryear{{Lee} \& {Saio}}{{Lee} \&
  {Saio}}{2020}]{lee20}
{Lee} U.,  {Saio} H.,  2020, \mn@doi [\mnras] {10.1093/mnras/staa2250}, \href
  {https://ui.adsabs.harvard.edu/abs/2020MNRAS.497.4117L} {497, 4117}

\bibitem[\protect\citeauthoryear{{Lewis}, {Bailey}, {Cotton}, {Howarth},
  {Kedziora-Chudczer}  \& {van Leeuwen}}{{Lewis} et~al.}{2022}]{lewis22}
{Lewis} F.,  {Bailey} J.,  {Cotton} D.~V.,  {Howarth} I.~D.,
  {Kedziora-Chudczer} L.,   {van Leeuwen} F.,  2022, \mn@doi [\mnras]
  {10.1093/mnras/stac991}, \href
  {https://ui.adsabs.harvard.edu/abs/2022MNRAS.513.1129L} {513, 1129}

\bibitem[\protect\citeauthoryear{{Lindegren} et~al.,}{{Lindegren}
  et~al.}{2021}]{Lindegren21}
{Lindegren} L.,  et~al., 2021, \mn@doi [\aap] {10.1051/0004-6361/202039653},
  \href {https://ui.adsabs.harvard.edu/abs/2021A&A...649A...4L} {649, A4}

\bibitem[\protect\citeauthoryear{{Marshall} et~al.,}{{Marshall}
  et~al.}{2016}]{marshall16}
{Marshall} J.~P.,  et~al., 2016, \mn@doi [\apj] {10.3847/0004-637X/825/2/124},
  \href {http://adsabs.harvard.edu/abs/2016ApJ...825..124M} {825, 124}

\bibitem[\protect\citeauthoryear{{Marshall}, {Cotton}, {Scicluna}, {Bailey},
  {Kedziora-Chudczer}  \& {Bott}}{{Marshall} et~al.}{2020}]{marshall20}
{Marshall} J.~P.,  {Cotton} D.~V.,  {Scicluna} P.,  {Bailey} J.,
  {Kedziora-Chudczer} L.,   {Bott} K.,  2020, \mn@doi [\mnras]
  {10.1093/mnras/staa3195}, \href
  {https://ui.adsabs.harvard.edu/abs/2020MNRAS.499.5915M} {499, 5915}

\bibitem[\protect\citeauthoryear{{Marshall}, {Cotton}, {Bott}, {Bailey},
  {Kedziora-Chudczer}  \& {Brown}}{{Marshall} et~al.}{2345}]{marshallPrep}
{Marshall} J.~P.,  {Cotton} D.~V.,  {Bott} K.,  {Bailey} J.,
  {Kedziora-Chudczer} L.,   {Brown} E.~L.,  \noop{2345}, {Multi-wavelength
  aperture polar\-imetry of debris-disc host stars}, in prep.

\bibitem[\protect\citeauthoryear{{Mason}, {Wycoff}, {Hartkopf}, {Douglass}  \&
  {Worley}}{{Mason} et~al.}{2001}]{mason01}
{Mason} B.~D.,  {Wycoff} G.~L.,  {Hartkopf} W.~I.,  {Douglass} G.~G.,
  {Worley} C.~E.,  2001, \mn@doi [\aj] {10.1086/323920}, \href
  {https://ui.adsabs.harvard.edu/abs/2001AJ....122.3466M} {122, 3466}

\bibitem[\protect\citeauthoryear{{Nu{\~n}ez} et~al.,}{{Nu{\~n}ez}
  et~al.}{2017}]{nunez17}
{Nu{\~n}ez} P.~D.,  et~al., 2017, \mn@doi [\aap] {10.1051/0004-6361/201730859},
  \href {https://ui.adsabs.harvard.edu/abs/2017A&A...608A.113N} {608, A113}

\bibitem[\protect\citeauthoryear{{Palmer}, {Walker}, {Jones}  \&
  {Wallis}}{{Palmer} et~al.}{1968}]{palmer68}
{Palmer} D.~R.,  {Walker} E.~N.,  {Jones} D.~H.~P.,   {Wallis} R.~E.,  1968,
  Royal Greenwich Observatory Bulletins, \href
  {https://ui.adsabs.harvard.edu/abs/1968RGOB..135..385P} {135, 385}

\bibitem[\protect\citeauthoryear{{Peterson} et~al.,}{{Peterson}
  et~al.}{2006}]{peterson06}
{Peterson} D.~M.,  et~al., 2006, \mn@doi [\apj] {10.1086/497981}, \href
  {https://ui.adsabs.harvard.edu/abs/2006ApJ...636.1087P} {636, 1087}

\bibitem[\protect\citeauthoryear{{Piirola} et~al.,}{{Piirola}
  et~al.}{2020}]{piirola20}
{Piirola} V.,  et~al., 2020, \mn@doi [\aap] {10.1051/0004-6361/201937324},
  \href {https://ui.adsabs.harvard.edu/abs/2020A&A...635A..46P} {635, A46}

\bibitem[\protect\citeauthoryear{{Reiners} \& {Royer}}{{Reiners} \&
  {Royer}}{2004}]{reiners04}
{Reiners} A.,  {Royer} F.,  2004, \mn@doi [\aap] {10.1051/0004-6361:20034175},
  \href {https://ui.adsabs.harvard.edu/abs/2004A&A...415..325R} {415, 325}

\bibitem[\protect\citeauthoryear{{Reiners} \& {Schmitt}}{{Reiners} \&
  {Schmitt}}{2002}]{reiners02}
{Reiners} A.,  {Schmitt} J.~H.~M.~M.,  2002, \mn@doi [\aap]
  {10.1051/0004-6361:20011801}, \href
  {https://ui.adsabs.harvard.edu/abs/2002A&A...384..155R} {384, 155}

\bibitem[\protect\citeauthoryear{{Reiners}, {Schmitt}  \&
  {K{\"u}rster}}{{Reiners} et~al.}{2001}]{reiners01}
{Reiners} A.,  {Schmitt} J.~H.~M.~M.,   {K{\"u}rster} M.,  2001, \mn@doi [\aap]
  {10.1051/0004-6361:20011023}, \href
  {https://ui.adsabs.harvard.edu/abs/2001A&A...376L..13R} {376, L13}

\bibitem[\protect\citeauthoryear{{Ricker} et~al.,}{{Ricker}
  et~al.}{2015}]{ricker15}
{Ricker} G.~R.,  et~al., 2015, \mn@doi [Journal of Astronomical Telescopes,
  Instruments, and Systems] {10.1117/1.JATIS.1.1.014003}, \href
  {https://ui.adsabs.harvard.edu/abs/2015JATIS...1a4003R} {1, 014003}

\bibitem[\protect\citeauthoryear{{Rieutord}, {Espinosa Lara}  \&
  {Putigny}}{{Rieutord} et~al.}{2016}]{rieutord16}
{Rieutord} M.,  {Espinosa Lara} F.,   {Putigny} B.,  2016, \mn@doi [Journal of
  Computational Physics] {10.1016/j.jcp.2016.05.011}, \href
  {https://ui.adsabs.harvard.edu/abs/2016JCoPh.318..277R} {318, 277}

\bibitem[\protect\citeauthoryear{{Royer}, {Grenier}, {Baylac}, {G{\'o}mez}  \&
  {Zorec}}{{Royer} et~al.}{2002}]{royer02}
{Royer} F.,  {Grenier} S.,  {Baylac} M.~O.,  {G{\'o}mez} A.~E.,   {Zorec} J.,
  2002, \mn@doi [\aap] {10.1051/0004-6361:20020943}, \href
  {https://ui.adsabs.harvard.edu/abs/2002A&A...393..897R} {393, 897}

\bibitem[\protect\citeauthoryear{{Saio}, {Kurtz}, {Murphy}, {Antoci}  \&
  {Lee}}{{Saio} et~al.}{2018}]{saio18}
{Saio} H.,  {Kurtz} D.~W.,  {Murphy} S.~J.,  {Antoci} V.~L.,   {Lee} U.,  2018,
  \mn@doi [\mnras] {10.1093/mnras/stx2962}, \href
  {https://ui.adsabs.harvard.edu/abs/2018MNRAS.474.2774S} {474, 2774}

\bibitem[\protect\citeauthoryear{{Seaton}}{{Seaton}}{1979}]{seaton79}
{Seaton} M.~J.,  1979, \mn@doi [\mnras] {10.1093/mnras/187.1.73P}, \href
  {https://ui.adsabs.harvard.edu/abs/1979MNRAS.187P..73S} {187, 73}

\bibitem[\protect\citeauthoryear{{Serkowski}}{{Serkowski}}{1958}]{serkowski58}
{Serkowski} K.,  1958, \actaa, \href
  {https://ui.adsabs.harvard.edu/abs/1958AcA.....8..135S} {8, 135}

\bibitem[\protect\citeauthoryear{{Serkowski}}{{Serkowski}}{1973}]{serkowski73}
{Serkowski} K.,  1973, in {Greenberg} J.~M.,  {van de Hulst} H.~C.,  eds,  IAU
  Symposium Vol. 52, Interstellar Dust and Related Topics. p.~145

\bibitem[\protect\citeauthoryear{{Serkowski}, {Mathewson}  \&
  {Ford}}{{Serkowski} et~al.}{1975}]{serkowski75}
{Serkowski} K.,  {Mathewson} D.~S.,   {Ford} V.~L.,  1975, \mn@doi [\apj]
  {10.1086/153410}, \href {http://adsabs.harvard.edu/abs/1975ApJ...196..261S}
  {196, 261}

\bibitem[\protect\citeauthoryear{{Slettebak}}{{Slettebak}}{1954}]{slettebak54}
{Slettebak} A.,  1954, \mn@doi [\apj] {10.1086/145804}, \href
  {https://ui.adsabs.harvard.edu/abs/1954ApJ...119..146S} {119, 146}

\bibitem[\protect\citeauthoryear{{Slettebak}}{{Slettebak}}{1966}]{slettebak66}
{Slettebak} A.,  1966, \mn@doi [\apj] {10.1086/148748}, \href
  {https://ui.adsabs.harvard.edu/abs/1966ApJ...145..126S} {145, 126}

\bibitem[\protect\citeauthoryear{{Slettebak}, {Collins}, {Boyce}, {White}  \&
  {Parkinson}}{{Slettebak} et~al.}{1975}]{slettebak75}
{Slettebak} A.,  {Collins} G.~W. I.,  {Boyce} P.~B.,  {White} N.~M.,
  {Parkinson} T.~D.,  1975, \mn@doi [\apjs] {10.1086/190338}, \href
  {https://ui.adsabs.harvard.edu/abs/1975ApJS...29..137S} {29, 137}

\bibitem[\protect\citeauthoryear{{Smith} \& {Gray}}{{Smith} \&
  {Gray}}{1976}]{smith76}
{Smith} M.~A.,  {Gray} D.~F.,  1976, \mn@doi [\pasp] {10.1086/130029}, \href
  {https://ui.adsabs.harvard.edu/abs/1976PASP...88..809S} {88, 809}

\bibitem[\protect\citeauthoryear{{Spurr}}{{Spurr}}{2006}]{spurr06}
{Spurr} R. J.~D.,  2006, \mn@doi [\jqsrt] {10.1016/j.jqsrt.2006.05.005}, \href
  {https://ui.adsabs.harvard.edu/abs/2006JQSRT.102..316S} {102, 316}

\bibitem[\protect\citeauthoryear{{Townsend}, {Owocki}  \& {Howarth}}{{Townsend}
  et~al.}{2004}]{townsend04}
{Townsend} R.~H.~D.,  {Owocki} S.~P.,   {Howarth} I.~D.,  2004, \mn@doi
  [\mnras] {10.1111/j.1365-2966.2004.07627.x}, \href
  {https://ui.adsabs.harvard.edu/abs/2004MNRAS.350..189T} {350, 189}

\bibitem[\protect\citeauthoryear{{Tsipouras} \& {Cormier}}{{Tsipouras} \&
  {Cormier}}{1973}]{tsipouras73}
{Tsipouras} P.,  {Cormier} R.,  1973, Technical Report~272, Hermite
  Interpolation Algorithm for Constructing Reasonable Analytic Curves through
  Discrete Data Points.
US Airforce Surveys in Geophysics, Cambridge, Mass.

\bibitem[\protect\citeauthoryear{{Uesugi} \& {Fukuda}}{{Uesugi} \&
  {Fukuda}}{1982}]{uesugi82}
{Uesugi} A.,  {Fukuda} I.,  1982, {Catalogue of stellar rotational velocities
  (revised)}.
Dept. of Astronomy, University of Kyoto

\bibitem[\protect\citeauthoryear{\VAN{Leeuwen}{van}{van}~Leeuwen}{\VAN{Leeuwen}{van}{van}~Leeuwen}{2007}]{vanLeeuwen07}
\VAN{Leeuwen}{van}{van}~Leeuwen F.,  2007, \mn@doi [\aap]
  {10.1051/0004-6361:20078357}, \href
  {https://ui.adsabs.harvard.edu/abs/2007A%26A...474..653V} {474, 653}

\bibitem[\protect\citeauthoryear{{Venn} \& {Lambert}}{{Venn} \&
  {Lambert}}{1990}]{venn90}
{Venn} K.~A.,  {Lambert} D.~L.,  1990, \mn@doi [\apj] {10.1086/169334}, \href
  {https://ui.adsabs.harvard.edu/abs/1990ApJ...363..234V} {363, 234}

\bibitem[\protect\citeauthoryear{{Wallenquist}}{{Wallenquist}}{1947}]{Wallenquist47}
{Wallenquist} A.,  1947, Uppsala Astronomical Observatory Annals, \href
  {https://ui.adsabs.harvard.edu/abs/1947UppAn...2b...1W} {2, 1}

\bibitem[\protect\citeauthoryear{{Wardle} \& {Kronberg}}{{Wardle} \&
  {Kronberg}}{1974}]{wardle74}
{Wardle} J.~F.~C.,  {Kronberg} P.~P.,  1974, \mn@doi [\apj] {10.1086/153240},
  \href {https://ui.adsabs.harvard.edu/abs/1974ApJ...194..249W} {194, 249}

\bibitem[\protect\citeauthoryear{{Westgate}}{{Westgate}}{1933}]{westgate33}
{Westgate} C.,  1933, \mn@doi [\apj] {10.1086/143483}, \href
  {https://ui.adsabs.harvard.edu/abs/1933ApJ....78...46W} {78, 46}

\bibitem[\protect\citeauthoryear{{Whittet}, {Martin}, {Hough}, {Rouse},
  {Bailey}  \& {Axon}}{{Whittet} et~al.}{1992}]{whittet92}
{Whittet} D.~C.~B.,  {Martin} P.~G.,  {Hough} J.~H.,  {Rouse} M.~F.,  {Bailey}
  J.~A.,   {Axon} D.~J.,  1992, \mn@doi [\apj] {10.1086/171039}, \href
  {https://ui.adsabs.harvard.edu/abs/1992ApJ...386..562W} {386, 562}

\bibitem[\protect\citeauthoryear{{Wilking}, {Lebofsky}, {Martin}, {Rieke}  \&
  {Kemp}}{{Wilking} et~al.}{1980}]{wilking80}
{Wilking} B.~A.,  {Lebofsky} M.~J.,  {Martin} P.~G.,  {Rieke} G.~H.,   {Kemp}
  J.~C.,  1980, \mn@doi [\apj] {10.1086/157694}, \href
  {https://ui.adsabs.harvard.edu/abs/1980ApJ...235..905W} {235, 905}

\bibitem[\protect\citeauthoryear{{Wu}, {Singh}, {Prugniel}, {Gupta}  \&
  {Koleva}}{{Wu} et~al.}{2011}]{wu11}
{Wu} Y.,  {Singh} H.~P.,  {Prugniel} P.,  {Gupta} R.,   {Koleva} M.,  2011,
  \mn@doi [\aap] {10.1051/0004-6361/201015014}, \href
  {https://ui.adsabs.harvard.edu/abs/2011A&A...525A..71W} {525, A71}

\bibitem[\protect\citeauthoryear{{Zechmeister} \& {K{\"u}rster}}{{Zechmeister}
  \& {K{\"u}rster}}{2009}]{zechmeister09}
{Zechmeister} M.,  {K{\"u}rster} M.,  2009, \mn@doi [\aap]
  {10.1051/0004-6361:200811296}, \href
  {https://ui.adsabs.harvard.edu/abs/2009A&A...496..577Z} {496, 577}

\makeatother
\end{thebibliography}



\appendix

\section{Summary of observing runs}
\label{sec:obsruns}

\begin{table*}
\caption{Summary of observing runs.}
\label{tab:runs}
\tabcolsep 3.5 pt 
\begin{tabular}{cl|ccrccccc|rr}
\toprule
\multicolumn{2}{c|}{} & \multicolumn{8}{c|}{Telescope and Instrument Set-Up$^a$}   &    \multicolumn{2}{c}{Tel.\ Calibration$^{b}$}    \\
Run ID & \multicolumn{1}{c|}{Date Range$^c$} & Instr. &  Tel.$^d$ & \multicolumn{1}{c}{f/} & Ap. & Mod. & Filter & Det.$^e$ & $n$ &   \multicolumn{1}{c}{$q_{\text{TP}}$} & \multicolumn{1}{c}{$u_{\rm TP}$} \\
 &  \multicolumn{1}{c|}{(UT)} &  &   &  & ($\arcsec$) &  &  &  &  &   \multicolumn{1}{c}{(ppm)} & \multicolumn{1}{c}{(ppm)} \\
\midrule
2005\_04 & 04/25--05/08   & PlanetPol     & WHT   &   11\phantom{*}   &  5.2 & PEM    & BRB           & APD   & 1 &  \multicolumn{2}{c}{(Note $f$)} \\
2015\_10 &  10/14--11/02   &   HIPPI       & AAT   &   8\phantom{*}    &  6.6 & BNS-E1 & $g^{\prime}$  & B     & 1 &  $-$50.4 $\pm$ 1.1 & $-$0.2 $\pm$ 1.1 \\
2017\_08 &  08/12--07/04   &   HIPPI       & AAT   &   8\phantom{*}    &  6.6 & BNS-E2 & 425SP         & B     & 1 &  $-$7.3 $\pm$ 3.6 &    8.5 $\pm$ 3.6 \\
& & & & & &                                                                                 & 500SP         & B     & 1 & $-$10.0 $\pm$ 1.7 & $-$0.4 $\pm$ 1.6 \\
& & & & & &                                                                                 & $g^{\prime}$  & B     & 1 &   $-$9.1 $\pm$ 1.5 & $-$2.6 $\pm$ 1.4 \\
& & & & & &                                                                                 & $r^{\prime}$  & R     & 1 &  $-$10.4 $\pm$ 1.3 & $-$7.0 $\pm$ 1.3 \\
& & & & & &                                                                                 & 650LP         & R     & 1 &  $-$8.2 $\pm$ 2.3 & $-$5.1 $\pm$ 2.4 \\
2018\_07&  07/15--07/23   & HIPPI-2       & AAT  &    16$^g$              & 11.9 & BNS-E4 & 425SP         & B     & 2 &    $-$5.6 $\pm$ 6.4 & 19.8 $\pm$ 6.3 \\
&                               &               &      &                    &      &        & 500SP         & B     & 2 &     1.9 $\pm$ 1.4 & 18.4 $\pm$ 1.4 \\
&                               &               &      &                    &      &        & $g^{\prime}$  & B     & 1 &  $-$12.8 $\pm$ 1.1 &  4.1 $\pm$ 1.0 \\
&                               &               &      &                    &      &        & $V$             & B     & 2 &  $-$20.3 $\pm$ 1.5 &  2.3 $\pm$ 1.5 \\
&                               &               &      &                    &      &        & $r^{\prime}$  & B     & 1 &  $-$10.4 $\pm$ 2.2 &  3.7 $\pm$ 2.2 \\
&                               &               &      &                    &      &        & $r^{\prime}$  & R     & 1 &  $-$12.7 $\pm$ 1.2 &  0.4 $\pm$ 1.2 \\
&                               &               &      &                    &      &        & 650LP         & R     & 1 &  $-$6.6 $\pm$ 1.9 &  4.0 $\pm$ 1.9 \\
\bottomrule
\end{tabular}
\begin{flushleft}
Notes:\newline 
\textbf{$^a$}  `Instr.' is the instrument; `Tel.', `f/' the telescope and its $f$-ratio; `; `Ap.' the angular diameter, on the sky, of the photo\-meter entrance aperture;  `Mod.' the modulator;   `Det.' the detector;  and $n$ the number of independent observations.
Further details, including transmission curves for all components and characterizations of each modulator at the relevant epochs, can be found in \citet{hough06} and  \citet{bailey20a}. \newline 
\textbf{$^b$} The observations used to determine the telescope-induced polarization, TP, and the high-polarization standards observed to calibrate position angle, are described by \citet[][run 2015\_10]{marshall16}, \citet[][run 2017\_08]{cotton19b}, and 
\citet[][run 2018\_07]{bailey20a}.\newline 
\textbf{$^c$} Dates given are inclusive of $\zeta$-Aql and standard-star observations. \newline 
\textbf{$^d$} WHT,  4.2-m William Herschel Telescope (altazimuth mount);  AAT, 3.9-m Anglo-Australian Telescope (equatorial mount).\newline
\textbf{$^e$} B, R indicate blue- and red-sensitive photo\-multiplier tubes, respectively (Section~\ref{sec:ppol}); APD indicates PlanetPol's Avalanche Photo-Diodes, which resulted in a broad red bandpass (BRB) extending beyond $\sim$1$\mu$m.  
\newline
\textbf{$^f$} The telescope-polarization function for this altazimuth telescope is discussed by \citet{bailey10}.\newline
\textbf{$^g$} Focal ratio increased by using a 2$\times$ negative achromatic lens.
\\
\end{flushleft}
\end{table*}

Technical details of the observing runs leading to the results given in Table~\ref{tbl:obs} are summarized in Table~\ref{tab:runs}.
HIPPI and HIPPI-2 are dual-beam photo\-polarimeters that use ferro-electric liquid crystals (FLCs) for primary modulation at 500~Hz, in order to overcome seeing noise
and thereby to achieve high precision.
 The FLC used for all HIPPI/-2 observations of $\zeta$~Aql was manufactured by Boulder Nonlinear Systems;  the performance of this unit has evolved over time, an issue addressed 
 by the reduction pipeline used to process all the data from both instruments \citep{bailey20a}.

We made use of two types of Hamamatsu photo\-multiplier tube (PMT) as detectors; the blue-sensitive H10720-210, which we denote `B', and red-sensitive H10720-20, `R'. Six broadband filters were employed: custom-built 425-nm and 500-nm `short-pass' and 650-nm `long-pass' filters (425SP, 500SP, and 650LP, respectively;  \citealt{bailey20a}), SDSS $g^\prime$ and $r^\prime$, and Johnson $V$. The $r^\prime$ filter was paired with both the R and B detectors; the 650LP filter only with R; and the remainder only with B. 

Telescope optics introduce a small, wavelength-dependent polarization,  which was removed by reference
to observations of low-polarization standard stars (cf.\ Table~\ref{tab:runs}).

\section{Rotational velocity}

\label{sec:rotv}

The equatorial rotation velocity provides an important constraint on the stellar mass.
A summary of published estimates of \vesini\ for \zAql\ is given in Table~\ref{tbl:vesini}; \citet{abt95} provide the only primary measure\-ment on the \citet{slettebak75} system that we have been able to locate. They note the absence of reliable fast-rotating calibrators in their dataset, and mark their measure\-ment as uncertain.   
We therefore undertook a new analysis
in order to determine a modern, precise value for \vesini, using the Fourier-transform method
(e.g., \citealt{Carroll33, smith76}).

\subsection{Data}

A search of on-line archives showed that observations obtained with the ESPaDOnS echelle spectropolarimeter at the Canada--France--Hawaii Telescope (CFHT) are of particularly good quality, and numerous.   
After rejecting a handful of relatively low-quality exposures, we corrected relevant sections of each of the remaining sixty-three spectra for
weak telluric absorption (dividing individual spectra by a scaled telluric template constructed from the data merged in topocentric velocity space), then merged them (after correcting to heliocentric velocities).    The resulting noise-weighted mean spectrum has a continuum signal:noise ratio of $\gtrsim$3000 at $\sim$635~nm (as measured from residuals to low-order polynomial fits), at a resolving power of $R\simeq65000$.    We identified no  obvious line-profile or radial-velocity variability in these data (Section~\ref{sec:notsb}).

\begin{table}
\caption{Literature \vesini\ values.}
\centering
\tabcolsep 6 pt
\begin{tabular}{lclc}
	\toprule
&Value (\kms)&$\quad$Source&Notes\\
	\midrule
\multicolumn{3}{l}{\textit{Primary sources:}}&\\
\T&175    & \citet{westgate33} &[1]\\
  &365    & \citet{slettebak54}\\
  &350    & \citet{slettebak66} &[2]\\
  &305    & \citet{palmer68}\\
\B&\phantom{:}295:   & \citet{abt95}\\
\multicolumn{3}{l}{\textit{Secondary sources:}}&\\
\T&335      & \citet{boyarchuk64}&[3]\\
  &345      & \citet{uesugi82}&[4]\\
\B&317      & \citet{royer02}&[5]\\
   \bottomrule
\end{tabular}
\begin{flushleft}
[1]  \zAqlx\ is identified by \citeauthor{westgate33} only as Boss~4858;  though unattributed, this refers to [Lewis] \citet{boss10}, 
not his son's later, better-known `General Catalogue' ([Benjamin] \citealt{boss37}).\newline
[2] Most tabulated values given to the nearest 50~\kms.\newline
[3] Appears to be a straight, albeit proleptic, average of \citet{slettebak54} and \citet{palmer68} values.\newline
[4] Weighted average of rescaled earlier results.\newline
[5] Rescaling of \citet{abt95} result.   Uncertainty of $\pm$38~\kms\ quoted by \citet{cochetti20}.

\end{flushleft}
\label{tbl:vesini}
\end{table}

\subsection{Modelling: ad hoc characterization of differential rotation}
\label{sec:lpm}

\begin{figure*}
\includegraphics[width=\textwidth]{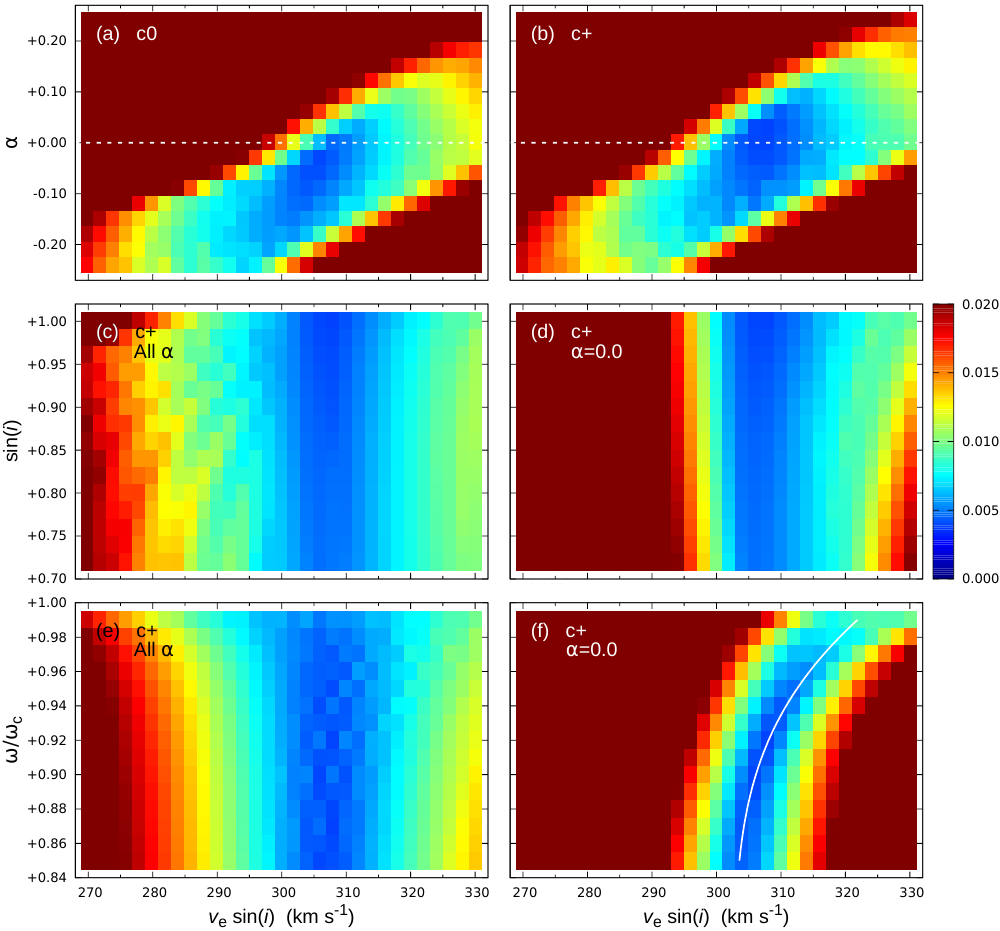}
\caption{Heatmaps summarizing comparisons of observed and model Fourier transforms for the \SiII~$\lambda$634.7 line;  the test statistic is the r.m.s. difference between observed and modelled normalized Fourier transforms (so smaller values mean better matches), taking the minimum values over marginal variables.   Panels~(a) and (b), marginalized 
over \omomc\ and $i$,
represent results for two slightly different rectifications of the observed profile, `c0' and `c+', described in
Section~\ref{sec:rotnum}. Panels~(c) and (e) are marginalized over all values of $\alpha$, while (d) and (f) are for solid-body surface rotation ($\alpha \equiv 0$;  panels (c)--(f) are all based on the c+ rectification).   The solid line in panel~(f) is eqtn.~\ref{eq:vom}.}
 \label{fig:rotmont}
\end{figure*}

Latitudinal differential rotation introduces changes to line-profile shapes, principally by modifying the Doppler redistribution of absorption arising at temperate latitudes (for given \ve\ and \sini;  changes in gravity darkening introduce further, but secondary, effects).  This in turn affects the Fourier transform of the profiles -- notably, the separation of the first and second minima (e.g., \citealt{reiners01, reiners02}).

Given the  quality of the CFHT data, and in the light of growing observational evidence for differential rotation in at least some A-type stars (e.g., \citealt{ammler12,balona16,kawaler21}), we chose to incorporate an empirical
investigation of the possibility of differential surface rotation into our initial analysis.
For these exploratory calculations, we characterized $\omega(\theta)$, the  angular rotation rate at colatitude $\theta$, by
\begin{equation}
\frac{\omega(\theta)}{\omega_{\rm e}} =
1 - \alpha + \alpha\left\{{
\frac{R(\theta)\,\sin(\theta)}{\req} }\right\}^2,
\label{eq:adhoc}
\end{equation}
where $\omega_{\rm e}$, \req\ are equatorial values;
this reduces to the de facto standard analytical form
${\omega(\theta)}/{\omega_{\rm e}} = 1- \alpha\cos^2(\theta)$ in the spherical-star limit.\footnote{We observe that this formulation, widely used in the cool-star community, is entirely ad hoc;   its form can be traced back to \citeauthor{carrington63} (\citeyear{carrington63}, p.~223, albeit with an exponent of \nicefrac{7}{4}).} The $\alpha$ parameter is positive for solar-type rotation (angular velocity greatest at the equator), with
$ \alpha_\odot \simeq +0.2$.

We computed synthetic spectra (incorporating full Roche-model rotational effects)
over a space intended to cover the likely range of parameter values at suitable sampling densities:

\hangindent=\parindent
\vesini\ in the range 270:330~\kms, at steps of 2~\kms;\\*
$\sini, 0$.72:1.00 @ 0.02;\\*
\hspace{5mm}$\omomc, 0$.85:0.99 @ 0.01; and \\*
$\alpha, -0.24$:+0.24 @ 0.03.

\hangindent=0cm\noindent
First results for regions around Ca\;{\sc ii}~$K$, Mg\;{\sc ii}~448.1~nm, and Si\;{\sc ii}~634.7~nm showed that significantly subsolar metallicity is required to match observed line strengths, in accord with reports by \citet{gray03} and \citet{wu11};  we obtained reasonable agreement for
[M/H] $\simeq -0.5$, and adopted that value.
We then focussed on the \SiII~$\lambda$634.7 line profile for analysis, as it is
one of the very few features not to show obvious blending in the spectra employed.   (Although least-squares deconvolution is ostensibly capable of addressing the blending issue, and of improving the overall signal:noise, its underpinning principles do not hold when gravity darkening is significant, as is the case here.  Moreover, systematics, rather than stochastic noise,  prove to dominate uncertainties in the conclusions.)  

For each model spectrum, the required values for \teff\ and polar radius (a surrogate for polar gravity in these circumstances) were those that reproduce the observed $V$, UV fluxes, for the matching \vesini, \omomc, and inclination values, as discussed in Section~\ref{sec:gin}, but for rigid rotation (regardless of the spectrum-synthesis value of~$\alpha$).   This approximation, adopted for computational expedience, is of no consequence for the \vesini\ analysis (as is also true for the adopted metallicity).

The comparison between observed and modelled $\lambda$634.7~nm profiles was conducted in Fourier space, using the r.m.s. differences between normalized transforms\footnote{`Normalized' here means dividing the power by the lowest-frequency value, which accounts for any small residual differences between observed and modelled line strengths.} as a test statistic.   Precise numerical results depend on the exact frequency (inverse velocity) interval chosen for the comparison, but our general conclusions are insensitive to this, for any reasonable values.   We used the range $\mbox{(1.5--4.5)} \times 10^{-3}$~km$^{-1}$~s, which encompasses the first two minima in the transform (Fig~\ref{fig:rotprof});   including the third minimum does not materially change any conclusions, but starts to run into the noise.     We found no evidence for any additional broadening processes (`macro\-turbulence') beyond the basic physical mechanisms integral to the modelling.

\begin{figure}
\includegraphics[width=\columnwidth]{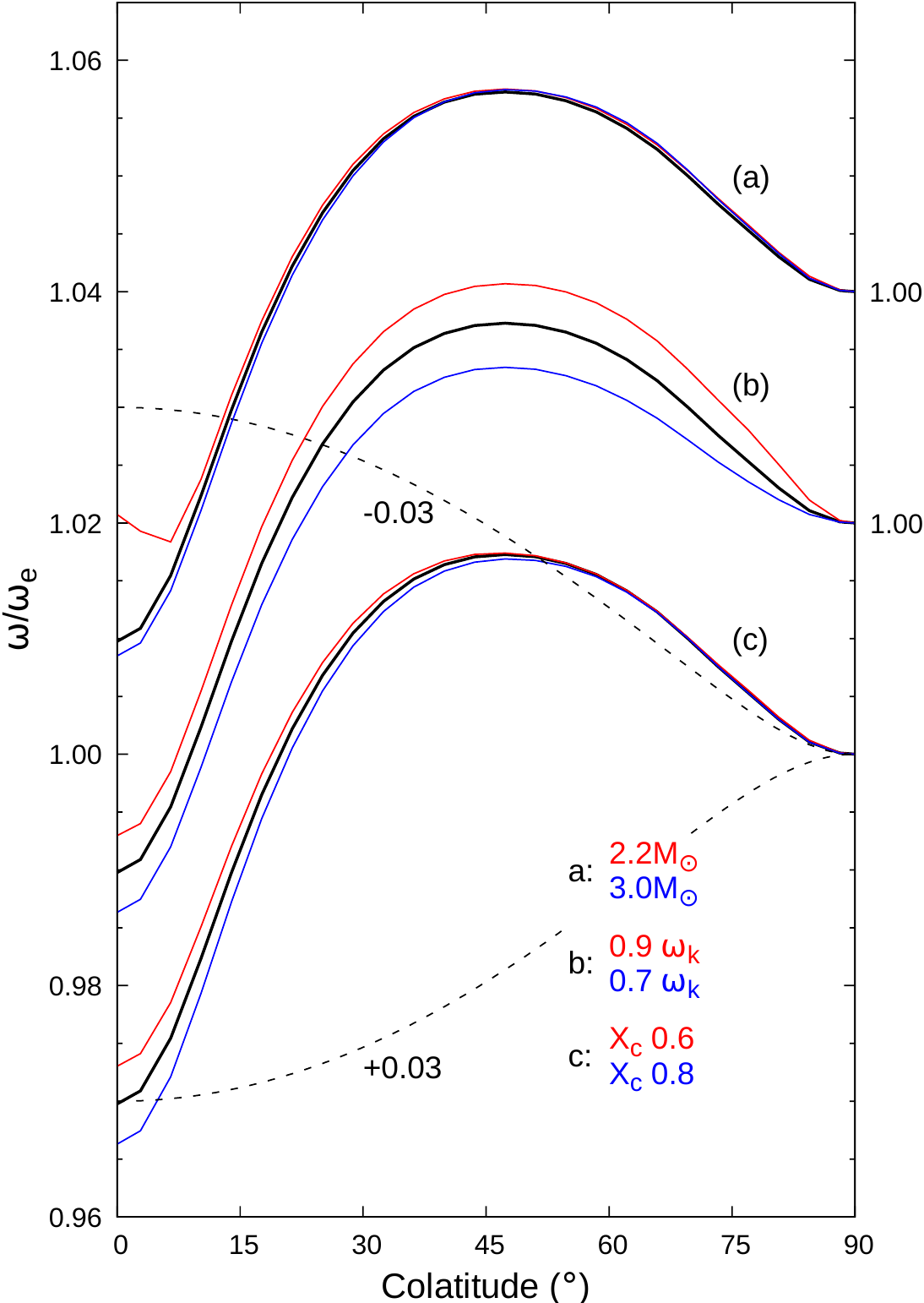}
\caption{Differential surface rotation predicted from \textsc{ester} \mbox{2-D} stellar-structure models.  The reference model in each group (shown in black) is for $M=2.5\msun$, $\xcore = 0.7$, $\omegae = 0.8\omegak$, with the sensitivity to these parameters illustrated by results for other values, as labelled. The dashed lines show the simple ad hoc differential-rotation characterization of eqtn.~\ref{eq:adhoc}, for the labelled values of the $\alpha$ parameter.}
\label{fig:ester}
\end{figure}

\begin{figure*}
\includegraphics[width=\textwidth]{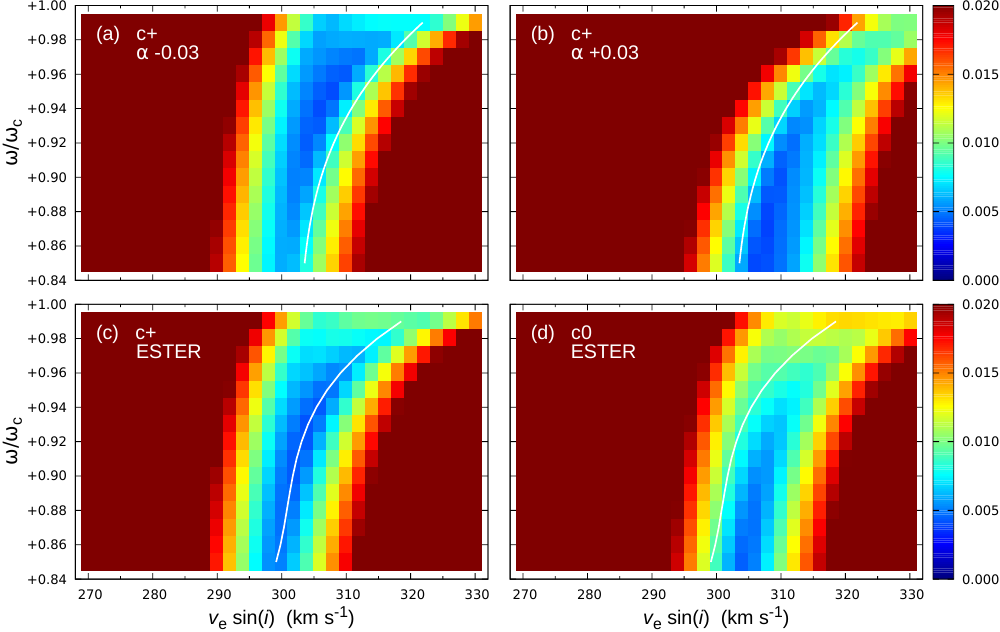}
\caption{The sensitivity of inferred \vesini\ to  assumptions in respect of
differential surface rotation and rectification.   Panels (a) and (b) are results for two different values of $\alpha$, assuming the ad hoc characterization of eqtn.~\ref{eq:adhoc};  for reference, the solid line is eqtn.~\ref{eq:voma}, the fit to the $\alpha=0.0$ (solid-body surface rotation) results shown in
Fig.~\ref{fig:rotmont}(f).   Panels (c) and (d) show results for the reference-model ESTER rotation profile (Fig.~\ref{fig:ester}), and two different rectifications discussed in Section~\ref{sec:rotnum}.}
\label{fig:rotdiff}
\end{figure*}

\subsection{Ad hoc characterization:  an empirical limit on differential rotation} 
\label{sec:rotnum}

Some results of the initial analysis are summarized in Fig.~\ref{fig:rotmont}.   The basic empirical test for differential rotation is embodied in panel~(a), where the minimum r.m.s. FT O$-$C for any (\sini, \omomc) combination is shown as a function of \vesini\ and $\alpha$.    This figure hints at possibly anti\-solar differential rotation (i.e., negative $\alpha$), which would contrast with the handful of positive-$\alpha$ A-star detections reported in the literature \citep{reiners04, ammler12}.  

However, we find that quite small revisions to the adopted continuum normalization can introduce significant changes to the transform (cf.\  \citealt{dravins90}).
We label our initial, `by eye', continuum as `c0' (Fig.~\ref{fig:rotmont}).
Modifying the observed profile by division [resp., multiplication] with a cosine bell of half-width 320~\kms\ and peak amplitude 0.1\%\ of this initial continuum  gives a slightly deeper [shallower] line of \textit{slightly} different shape arising from the slightly higher [lower] continuum, labelled c+ [c$-$].   The c+ continuum leads to the results shown in   Fig.~\ref{fig:rotmont}(b), which are entirely consistent with solid-body surface rotation.     

Continuum uncertainties in the rectified observations are certainly possible at this level (if only because of unrecognized weak line blends).  Furthermore, although the comparison model spectra can be rectified simply by division with the corresponding model continuum, in practice this does not lead to a result well suited to comparison to observations.  A degree of subjectivity therefore also enters in rectifying the models (even though this was done in an automated procedure), accommodating further potential uncertainty.
We conclude that a conservative interpretation of our results is that the initial line-profile analysis alone does not provide any compelling, direct evidence for differential rotation in \zAql, and constrains $|\alpha|$ to $\lesssim 0.05$.

\subsection{Ad hoc characterization:  \texorpdfstring{\omomc}{omegafactor} dependence} 
\label{sec:rotnumx}

Fig.~\ref{fig:rotmont} also illustrates the sensitivity of \vesini\ to other parameters of interest.   The line profile offers no useful diagnostic potential for axial inclination, but there is a clear dependence of \vesini\ on \omomc\ (at any fixed $\alpha$; e.g., panel~f).   This has a straightforward interpretation: as a consequence of gravity darkening, the high-velocity equatorial belt becomes less evident in the spectrum at high \omomc, requiring an increase in \vesini\ in order to fill in the extreme wings of the line profile (cf.\ \citealt{townsend04}).\footnote{In this particular case, the line equivalent width is roughly constant over the range of relevant temperatures, and it is the temperature dependence of the \textit{continuum} that is the dominant effect.   The `visible' parts of the star still provide sufficient information to constrain \mbox{\vesini} and $\alpha$, for given \omomc.}

To characterize this dependency, we estimated the \vesini\ value that gives the smallest r.m.s. at each sampled value of \omomc\
by using a Hermite interpolation formula \citep{tsipouras73,hill82}, and made polynomial fits to the results to obtain approximate analytical representations:
\begin{subequations}\label{eq:vom}
\begin{align}
{\vesini} = 
  &306.10 + \varpi\times(68.0 + \varpi\times(740 - 3709\varpi))\label{eq:voma}\\
  &306.10 + \varpi\times(82.4 + \varpi\times(769 - 3076\varpi))\label{eq:vomb}
\end{align}
\end{subequations}
(in \kms), where $\varpi = \omomc - 0.9$, and the (a), (b) numerical values are from 
otherwise identical analyses based on \hipparcos\ and \gaia\ parallaxes, respectively (and confirm the expectation of negligible sensitivity of \vesini\ to distance).
Equation~\ref{eq:voma} is
shown as a white line in Fig.~\ref{fig:rotmont}(f), and
represents our adopted characterization of \vesini\ as a function of \omomc\ for these 
$\alpha \equiv 0$ models (valid over the  range $0.85\le\omomc\lesssim0.99$).  

\subsection{ESTER modelling}
\label{sec:estermod}

As discussed in Section~\ref{sec:pprob}, an initial analysis based on assumed solid-body surface rotation and the \hipparcos\ parallax was challenged by  disparities between rotation periods implied by the models and the \textit{TESS} photo\-metric period. We therefore  examined  the  question of differential rotational further, under the constraint of theoretical models of differential surface rotation, rather than an arbitrary ad hoc formulation.   To this end we computed a series of structure models using the \ester\ code\footnote{\texttt{http://ester-project.github.io/ester/}} (\citealt{espinosa13}; \citealt{rieutord16}).  \Ester\ 
computes the stellar structure self-consistently with the radial and latitudinal differential rotation  and the meridional circulation resulting from driving by the baroclinic torque (at solar abundance).

The models depend principally on three parameters:  mass, relative core-hydrogen abundance ($\xcore$, a surrogate for evolutionary stage;  $\xcore = 1\rightarrow0$, $\text{ZAMS}\rightarrow\text{TAMS}$), and the equatorial angular velocity, conventionally expressed in this context with respect to the Keplerian value,
\begin{equation}
\omegak = \sqrt{GM/\req^3},\; = \omcrit (1.5\rpole/\req)^{3/2}.
\end{equation}
For reference, $\omegae/\omegak = 0.7, 0.8, 0.9$ corresponds to $\omomc = 0.93, 0.97, 0.99$, a range relevant to our results for \zAql.

Initial parameter modelling  indicated \mbox{$M \simeq 2.5\msun$}, \mbox{$\xcore \simeq 0.7$}, \mbox{$\omegak \simeq 0.8$}; \ester\ surface-rotation results for this parameter set are shown in Fig.~\ref{fig:ester} (and were used for most of our subsequent modelling).   The sensitivity to parameter variations from this baseline set is also illustrated.   Even at rather rapid rotation, only quite modest departures from solid-body surface rotation are predicted, and they are insensitive to precise values of the free parameters (within the range of uncertainty of our results).    

As concluded in Section~\ref{sec:rotnum}, such modest departures from solid-body surface rotation are not directly detectable through our empirical line-profile analysis (they produce results corresponding, very roughly, to $\alpha \simeq -0.03$).   Nevertheless, they do introduce small but significant changes to the inferred \vesini\ values.   We therefore repeated the analysis of Section~\ref{sec:lpm}, incorporating the baseline \ester\ differential-rotation profile. Selected results are included in Fig.~\ref{fig:rotdiff}.   An analytical approximation to this additional set of \vesini\ vs.\ \omomc\ results, plotted in Fig.~\ref{fig:rotdiff}(c), is 
\begin{align}
\label{eq:vomester}
{\vesini} = 
 301.81 +
         \varpi\times(38.7 + \varpi\times(397 + 16735\varpi))
\end{align}
(\ester\ rotation profile, \gaia\ parallax, `c+' continuum);   that is, the \ester\ rotation profile leads to inferred \vesini\ values that are $\sim$4~\kms, or $\sim$1\%, smaller than solid-body rotation.   This is evidently a consequence of averaging over the equatorial and super-rotating
temperate latitudes.

\bsp	
\label{lastpage}
\end{document}